\documentclass[%
reprint, amsmath,amssymb, aps, nofootinbib]{revtex4-2}

\usepackage{feynmp-auto}
\usepackage{slashed}
\usepackage{amsmath,amsfonts,amssymb,amsthm,bm}
\usepackage{graphicx}
\usepackage[hidelinks]{hyperref}
\usepackage{color}

\newcommand{\deriv}[2]{\frac{\,d #1}{\,d #2}}
\newcommand{\doublederiv}[3]{\frac{\,d^2 #1}{\,d #2 \,d #3}}

\definecolor{turq}{rgb}{.1,.3,.5}
\definecolor{cela}{rgb}{.0,.6,.5}

\usepackage{placeins}

\begin{document}

\preprint{APS/123-QED}

%\title{Approximations for first-order QED processes in Monte Carlo simulations of laser and pulsar plasmas}
%\title{Efficient approximations of first-order QED processes for Monte Carlo simulation of laser and pulsar plasmas}
\title{Monte Carlo sampling of first-order QED processes in laser and pulsar plasmas}

\author{Verneri Sarjomaa}
    \email{verneri.sarjomaa@helsinki.fi}
\affiliation{Department of Physics, University of Helsinki, P.O. Box 64, FI-00014 University of Helsinki, Finland}
\author{Joonas Nättilä}
 \email{joonas.nattila@helsinki.fi}
\affiliation{Department of Physics, University of Helsinki, P.O. Box 64, FI-00014 University of Helsinki, Finland}

\date{\today}

\begin{abstract}
Monte Carlo sampling of strong-field quantum electrodynamics processes underpins simulations of high-intensity laser experiments and of astrophysical compact-object magnetospheres. Sampling an event requires the total rate of the process together with the cumulative probability that determines how energy is partitioned between the produced particles. Simulations typically tabulate both in advance and invert the tabulated probability numerically. Here we replace this procedure with elementary-function approximations for synchrotron radiation and the nonlinear Breit--Wheeler process. For each process, we approximate the auxiliary function that sets the total rate, as well as the cumulative probability, with Pad\'e approximants chosen so that the inversion reduces to a quartic equation. This yields the sampled quantum parameter---electron $\chi_e$ or photon $\chi_\gamma$---in closed form. The approximations and the particle spectra sampled from them agree with the exact results to within $1\%$. The procedure requires no lookup tables, no interpolation, and no numerical root finding, and can be inserted directly into radiative particle-in-cell codes.
\
\end{abstract}

\maketitle

\section{Introduction}\label{sect:intro}

Simulations of quantum electrodynamical (QED) processes in strong electromagnetic fields have become an active field of research, driven by applications in laser physics and astrophysics.
Upcoming high-intensity laser facilities will reach intensities beyond $10^{22}~\mathrm{W/cm}^2$, where QED processes start to govern the dynamics of the irradiated plasma, and simulations are the main tool for predicting what those experiments will see \cite{grismayer2017seeded}.
Understanding this regime matters both for the development of the lasers themselves and for their interaction with matter \cite{nerush2011laser}.
In astrophysics, the same processes drive $e^\pm$-pair cascades in neutron star magnetospheres \cite{timokhin2010time} and shape the radiative dynamics of black hole plasmas \cite{nattila2024radiative}.

Simulations account for the indeterministic nature of these processes with Monte Carlo (MC) sampling \cite{lobet2016modeling, niel2018quantum, kirk2014modelling, duclous2011monte}.
The algorithm draws two random numbers per event:
the first fixes the time a particle propagates in the field before the process occurs, and the second fixes how the energy is split between the produced particles.
The first draw needs the total rate of the process, which is set by an auxiliary function $T(\chi)$ of the quantum parameter $\chi$;
the second needs the cumulative probability $p(\chi,\chi_1)$ built from the differential spectrum, inverted for the   quantum parameter $\chi_1$ of the produced particle.
Both quantities contain integrals of modified Bessel functions, which are far too slow to evaluate at every time step of every particle.
Simulations therefore evaluate them in advance and store them in lookup tables, and previous work on the efficiency of MC sampling has concentrated on choosing optimal tabulations and interpolation schemes \cite{volokitin2023optimized, guo2022improving}.

We take a different route and replace the tabulated functions by approximations written in elementary functions alone.
We construct such approximations for the auxiliary functions and the cumulative probabilities of synchrotron radiation and of the nonlinear Breit-Wheeler process, hereafter the Breit-Wheeler process.
The cumulative probabilities are approximated by fourth-order Padé approximants in the energy fraction $r = \chi_1/\chi$.
This form is flexible enough to follow the exact probability across the full range of quantum parameters and, more importantly, it is invertible in closed form:
the inversion reduces to a quartic equation whose relevant root we give explicitly.
The quantum parameter of the produced particle then follows from the random number by direct evaluation, with no table, no interpolation, and no iterative root finding.
The approximations are also independent of any particular tabulation, so they can be used with any input values inside their region of validity.

We fit the approximations by weighted error minimization against numerically computed values, and we test them in two ways:
by comparing the derivative of the approximated probability with the exact spectrum, and by sampling events and comparing the resulting histograms with the theoretical distributions.
The approximations of the auxiliary functions stay within $0.7\%$ (synchrotron) and $0.1\%$ (Breit-Wheeler, for $\chi_\gamma \gtrsim 1$) of the exact functions, and the sampled spectra agree with theory at the percent level.

\section{Theory}\label{sect:theory}

\begin{figure}[t]
\centering
\includegraphics[width=\columnwidth, clip=true, trim={5.5cm 21cm 5.5cm 1.85cm}]{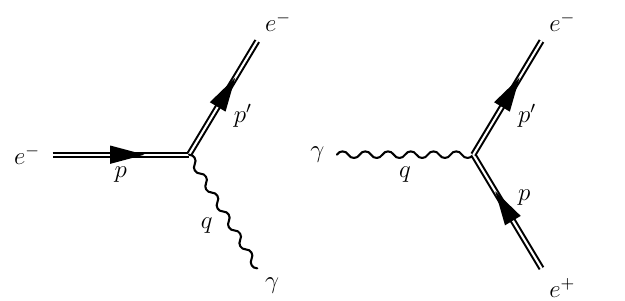}
\caption{Feynman diagrams of the two first-order processes. Synchrotron radiation (left): an electron of momentum $p$ emits a photon of momentum $q$ and continues with momentum $p'$. Breit-Wheeler pair production (right): a photon of momentum $q$ decays into an electron and a positron. The double lines denote fermion states dressed by the background field.}
\label{fig:feynman}
\end{figure}

QED in an external electromagnetic field $\mathcal{A}$ is governed by the Lagrangian $\mathcal{L} = \mathcal{L}_0+\mathcal{L}_\mathrm{int}$ (in natural units $\hbar=c=1$), whose free part is
\begin{equation}
  \mathcal{L}_0=-\frac{1}{4}(F_{\mu \nu})^2 + \overline{\psi}(i\slashed{\partial} -e\slashed{\mathcal{A}} - m)\psi,
\end{equation}
and whose electromagnetic interactions sit in $\mathcal{L}_\mathrm{int}=-e\overline{\psi}\slashed{A} \psi$.
Here $\slashed{a} \equiv \gamma^\mu a_\mu$ denotes a four-vector contracted with the gamma matrices, and $F_{\mu \nu}=\partial_\mu A_\nu - \partial_\nu A_\mu$ is the electromagnetic field strength tensor \cite{advancesinqed}.
The interaction part generates the QED vertex factor $-ie\gamma^\mu$, to be integrated over the vertex position $x$.
We study the processes in an evolving plane-wave background that depends only on the phase $\phi$,
\begin{equation} \label{eq:background_field}
  \mathcal{A}^\mu(\phi)=\mathcal{A}_0\phi \varepsilon_\mathcal{A}^\mu,
\end{equation}
with wave phase $\phi=k\cdot x$ and polarization vector $\varepsilon_\mathcal{A}=(0,1,0,0)$.
We choose the coordinate system so that the wave travels in the negative $z$-direction, $k=\omega(1,0,0,-1)$, which makes the momentum light-like and puts the wave in the Lorentz gauge, $k^2=0$ and $k_\mu \varepsilon_\mathcal{A}^\mu=0$.
This choice of polarization corresponds to a constant crossed electromagnetic field
\begin{align}
  \mathbf{E} &=-\mathcal{A}_0\omega\hat{x}, \\
  \mathbf{B} &=\mathcal{A}_0\omega\hat{y}. \label{eq:const_b}
\end{align}
Fermions are described by the solutions of the Dirac equation with the external background field included, $(i\slashed{\partial} - e\slashed{\mathcal{A}} - m)\psi_{p,r}(x)=0$.
For phase-dependent backgrounds, the ansatz $\psi=f(\phi)e^{-ipx}u_{p,r}$ yields the Volkov states, where $u_{p,r}$ is a Dirac spinor of spin state $r$ and momentum $p$.
An incoming fermion is described by the Volkov state \cite{seipt2017volkovstatesnonlinearcompton}
\begin{equation}
  \psi_{p,r}(x) = \left(1+\frac{e\slashed{k}\slashed{\mathcal{A}}}{2k \cdot p}\right)e^{-ipx-i\Phi_p(\phi)}u_{p,r},
\end{equation}
where the function in the complex exponential is
\begin{equation}
  \Phi_p(\phi)=\int_{0}^\phi \frac{d\phi'}{2k\cdot p}\left[2e\mathcal{A}(\phi')\cdot p - e^2 \mathcal{A}^2(\phi')\right].
\end{equation}
An outgoing fermion is described by the Dirac adjoint $\overline{\psi}=\psi^\dagger \gamma^0$,
\begin{equation}
  \overline\psi_{p,r}(x)=\overline u_{p,r}\left(1+\frac{e\slashed{\mathcal{A}}\slashed{k}}{2k\cdot p}\right)e^{ipx+i\Phi_p(\phi)}.
\end{equation}
An incoming photon of momentum $q$ is described by $A_\mu(x) =\varepsilon_\mu e^{-iqx}$, where $\varepsilon$ is the polarization vector, and an outgoing photon by $A^*_\mu(x) = \varepsilon^*_\mu e^{iqx}$.

The processes are defined entirely by the Lorentz-invariant $\chi$-parameters of the participating particles.
The parameter follows from the external field strength tensor $\mathcal{F}_{\mu \nu}=\partial_\mu \mathcal{A}_\nu - \partial_\nu \mathcal{A}_\mu$.
For a massive particle of mass $m$, charge $q$, and four-momentum $p^\nu$ it is \cite{lobet2016modeling}
\begin{equation}
    \chi =  \frac{\hbar}{m^3 c^3}  |q\mathcal{F}_{\mu \nu}p^\nu|  \, ,
\end{equation}
and for photons of momentum $p^\nu = \hbar k^\nu$
\begin{equation} \label{eq:chi_photon}
  \chi_\gamma = \frac{\hbar^2 e}{m^3_ec^3} |\mathcal{F}_{\mu \nu}k^\nu|,
\end{equation}
where $e$ is the elementary charge and $m_e$ is the electron mass.
In the ultra-relativistic limit the quantum parameters and the normalized energies are connected via $\chi_\gamma \gamma_e \approx \chi_e \gamma_\gamma$ \cite{niel2018quantum}, where $\gamma_e=\varepsilon_e/(m_ec^2)$ and $\gamma_\gamma=\varepsilon_\gamma/(m_ec^2)$ are set by the electron and photon energies $\varepsilon_e$ and $\varepsilon_\gamma$.
Kinematics conserves the sum of the $\chi$-parameters, $\sum_k \chi^\mathrm{in}_{k} = \sum_k \chi^\mathrm{out}_{k}$.
Particles with $\chi \ll 1$ behave classically, whereas particles with $\chi \gtrsim1$ are in the regime where quantum effects dominate.

\subsection{Calculation of the process spectra}

Feynman diagrams give the matrix elements from which the rates and spectra follow.
Here we display the main steps for the photon emission spectrum of synchrotron radiation, and then obtain the Breit-Wheeler spectrum by crossing symmetry, that is, by exchanging external momenta.
The Feynman rules give the matrix element of the synchrotron radiation process \cite{seipt2017volkovstatesnonlinearcompton, nikishov1964quantum, ritus1985quantum}
\begin{equation}
  S= -ie\int d^4 x\overline{\psi}_{p'}(x)\gamma^\mu \varepsilon^*_\mu\psi_{p}(x) e^{iqx}.
\end{equation}
The position dependence outside the exponentials originates from the phase dependence of the background field.
All other spatial dependence sits inside exponentials, which yield delta functions upon integration.
Phase-dependent terms are most naturally handled in light-front (LF) coordinates, which carry the phase as one of their dimensions;
App.~\ref{app:lfcoord} collects their properties.
Integrating over $x^-$ and $\mathbf{x}^\perp$ in LF coordinates converts the matrix element to
\begin{equation}
  S = -(2\pi)^4 ie\int_{-\infty}^\infty dn\overline{u}_{p', r} \Gamma u_{p,s} \delta(nk + p - p' - q),
\end{equation}
where the structure factor is
\begin{equation}
\begin{split}
    \Gamma = \slashed{\varepsilon}^* F&-\frac{ie\mathcal{A}_0}{2}\left(\frac{\slashed{\varepsilon}^* \slashed{k}\slashed{\varepsilon}_A}{k\cdot p} + \frac{\slashed{\varepsilon}_A \slashed{k} \slashed{\varepsilon}^*}{k\cdot p'}\right)F'\\ &+\frac{e^2 \mathcal{A}_0^2(\varepsilon_\mathcal{A})^2}{2(k\cdot p')(k\cdot p)}(\varepsilon^* \cdot k) \slashed{k} F''.
\end{split}
\end{equation}
The function $F$ and its derivatives are defined through the Airy function $\mathrm{Ai}(y)$ as
\begin{equation}
  F(n, \alpha, \beta)=(4\beta)^{-1/3}\exp{\left(\frac{i\alpha n}{8\beta}-\frac{i\alpha^3}{192\beta^2}\right)}\mathrm{Ai}(y),
\end{equation}
with the argument $y= (4\beta)^{2/3}[n/(4\beta)-(\alpha/8\beta)^2]$ and parameters
\begin{align}
  \alpha &=e\mathcal{A}_0\left(\frac{\varepsilon_\mathcal{A} \cdot p}{k\cdot p}- \frac{\varepsilon_\mathcal{A} \cdot p'}{k\cdot p'}\right), \\
  \beta  &=\frac{e^2\mathcal{A}_0^2\varepsilon_\mathcal{A}^2}{8}\left(\frac{1}{k\cdot p}- \frac{1}{k\cdot p'}\right).
\end{align}
The Airy functions appear when the phase integrals are absorbed into
\begin{equation}
\begin{split}
    &\int d\phi \phi^m e^{i\left(-\frac{\alpha}{2}\phi^2+\frac{4\beta}{3}\phi^3+n'\phi\right)}=\\ &2\pi(-i)^m \int_{-\infty}^\infty dn\delta(n'-n)\frac{\partial^m}{\partial n^m}F(n, \alpha, \beta).
\end{split}
\end{equation}
The new integration runs over the parameter $n$, which is the fraction of background field momentum exchanged in the process.

The rates follow by squaring the matrix element, summing over final spin and photon polarizations, and averaging over the initial spin,
\begin{equation}
  \frac{1}{2}\sum_{\substack{r,s\\ \lambda}}\frac{|S|^2}{TV}\equiv\int_{-\infty}^\infty dn|\mathcal{M}|^2 \delta(nk+p-p'-q),
\end{equation}
where $\chi_{e'}$ is the quantum parameter of the outgoing electron, $\chi_{e'}=\chi_e-\chi_\gamma$, and

\begin{align}
\begin{split} &|\mathcal{M}|^2=\frac{8(2\pi)^5m_e^2e^2}{a_0^2L_\phi }\left(\frac{\chi_\gamma}{2\chi_e\chi_{e'}}\right)^{-\frac{2}{3}}\bigg[-\mathrm{Ai}(y)^2\\&+ \left(\frac{\chi_\gamma}{2\chi_e\chi_{e'}}\right)^{-\frac{2}{3}}\left(1+\frac{\chi_\gamma^2}{2\chi_e\chi_{e'}}\right)\left(y\mathrm{Ai}(y)^2 + \mathrm{Ai}'(y)^2\right)\bigg].
\end{split}
\end{align}
Here $a_0=e\mathcal{A}_0/m$ is the normalized field strength and $L_\phi$ is the interval spanned by the phase of the background field.
The latter comes from the volume of the light-front coordinates and cancels against the integration over photon momentum.
Integrating over the phase space gives the number of photons emitted per unit time \cite{berestetskii2012quantum}
\begin{equation}
  \frac{dN_\mathrm{S}}{dt} = \int_{-\infty}^\infty dn\frac{|\mathcal{M}|^2}{2p_0}\frac{d^3 p' d^3q}{(2\pi)^3 2p'_0(2\pi)^3 2q_0}\delta(nk + p - p' - q).
\end{equation}
Integrating away the outgoing electron momentum and $q^1$ leaves the differential emission rate spectrum
\begin{equation}
\begin{split}
    &\doublederiv{N_\mathrm{S}}{\chi_\gamma}{t}=-\frac{m_e^2 e^2}{4\pi \varepsilon_e \chi_e}\bigg\{\int_{\left(\frac{\chi_\gamma}{\chi_e\chi_{e'}}\right)^{2/3}} ^\infty \mathrm{Ai}(z)dz  \\&+\left(\frac{\chi_\gamma}{\chi_e\chi_{e'}}\right)^{-2/3}\left(2+\frac{\chi_\gamma^2}{\chi_e\chi_{e'}}\right)\mathrm{Ai}'\left[\left(\frac{\chi_\gamma}{\chi_e\chi_{e'}}\right)^{2/3}\right]\bigg\},
\end{split}
\end{equation}
where we changed the integration over $q^3$ to one over $q^+$ (see App.~\ref{app:lfcoord}) and then to one over $\chi_\gamma$ using Eq.~\eqref{eq:chi_photon}.
Replacing the Airy functions by modified Bessel functions of the second kind \cite{abramowitz1948handbook}, $\mathrm{Ai}(x)=\sqrt{x/3\pi^2}\,K_{1/3}[(2/3)x^{3/2}]$ and $\mathrm{Ai}'(x)=-(x/\sqrt{3\pi^2})\,K_{2/3}[(2/3)x^{3/2}]$, casts the spectrum in its conventional form,
\begin{equation}
\begin{split}
    \doublederiv{N_\mathrm{S}}{\chi_\gamma}{t}=-\frac{\alpha_f m_e^2}{\pi \sqrt{3}\varepsilon_e \chi_e}\bigg[&\int_{\xi}^{\infty}K_{1/3}(s)\,ds \\&-\left(2+\frac{3\chi_\gamma\xi}{2}\right)K_{2/3}(\xi)\bigg],
\end{split}
\end{equation}
where $\xi=2\chi_\gamma/[3\chi_e(\chi_e-\chi_\gamma)]$.
Converting to SI units with $m_e\to m_ec^2$ and $\varepsilon\to \hbar \varepsilon$ gives
\begin{equation}
\begin{split} \label{eq:dn_s}
    \doublederiv{N_\mathrm{S}}{\chi_\gamma}{t}=-\frac{\alpha_f m_e^2c^4}{\pi \sqrt{3}\hbar \varepsilon_e \chi_e}\bigg[&\int_{\xi}^{\infty}K_{1/3}(s)\,ds \\&-\left(2+\frac{3\chi_\gamma\xi}{2}\right)K_{2/3}(\xi)\bigg],
\end{split}
\end{equation}
with the fine structure constant $\alpha_f=e^2/4\pi \approx 1/137$.

The Breit-Wheeler spectrum follows by interchanging the momenta of the external particles.
Figure~\ref{fig:feynman} shows that the momenta change as $q\to -q$, $p'\to p'$, and $p\to -p$ \cite{peskin2018introduction}.
Exchanging in- and outgoing particles also replaces $\varepsilon^*_\mu \to \varepsilon_\mu$ and $u_{p,r} \to v_{p,r}$.
The spin and polarization sums and the phase-space integration change accordingly, which replaces $\varepsilon_e \chi_e\to \varepsilon_\gamma\chi_\gamma$ in the spectrum, and an overall minus sign appears because a fermion moves from the initial to the final state.
The resulting pair-production spectrum is
\begin{equation}
\begin{split} \label{eq:dn_bw}
    \frac{d^2N_\mathrm{BW}}{d\chi_e dt}=\frac{\alpha_f m_e^2 c^4}{\pi \sqrt{3} \hbar \varepsilon_\gamma \chi_{\gamma} }\bigg[&\int_{\xi} ^\infty K_{1/3}(s)ds\\&- \left(2-\frac{3\chi_\gamma \xi}{2}\right)K_{2/3}(\xi)\bigg],
\end{split}
\end{equation}
where now $\xi = 2\chi_\gamma/[3\chi_e(\chi_\gamma-\chi_e)]$.
The Breit-Wheeler spectrum can be written with $\chi_e$ referring to either the electron or the positron.
We use it for the electron throughout and write the positron quantum parameter as $\chi_{e^+}$.

For both processes, the total rate follows by integrating the differential spectrum over all quantum parameter values available to the produced particle, $dN/dt=\int_0^{\chi}(d^2N/d\chi' dt)\,d\chi'$, and both rates can be expressed through an auxiliary function $T(\chi)$ as
\begin{equation}
  \frac{dN}{dt}=\frac{\alpha_f m_e^2c^4}{\hbar \varepsilon}\chi T(\chi),
\end{equation}
where $\varepsilon$ and $\chi$ are the energy and quantum parameter of the incoming particle.

\subsection{Synchrotron radiation}

The photon emission spectrum defines the radiation power spectrum
\begin{equation} \label{eq:prad}
  \frac{d P_\mathrm{rad}}{d\chi_\gamma}=\varepsilon_\gamma \frac{d^2N_\mathrm{S}}{d\chi_\gamma dt}
                                       = \frac{\varepsilon_e\chi_\gamma}{\chi_e}\cdot\frac{d^2N_\mathrm{S}}{d\chi_\gamma dt}.
\end{equation}
An electron of quantum parameter $\chi_e$ emits photons at the total rate
\begin{equation} \label{eq:rate_sync}
  \frac{dN_\mathrm{S}}{dt}=\frac{\alpha_f m_e^2 c^4}{\hbar \varepsilon_e} \chi_e T_\mathrm{S}(\chi_e),
\end{equation}
where the auxiliary function is
\begin{equation} \label{eq:t_sync_def}
\begin{split}
    T_\mathrm{S}(\chi_e) = -\frac{1}{\pi \sqrt{3} \chi^2_e}\int_0^{\chi_e}\bigg[&\int_{\xi}^{\infty}K_{1/3}(s)\,ds\\&-\left(2+\frac{3\chi_\gamma \xi}{2}\right)K_{2/3}(\xi)\bigg] d\chi_\gamma,
\end{split}
\end{equation}
with the same $\xi$-parameter as in Eq.~\eqref{eq:dn_s}.
The asymptotic forms of the auxiliary function are \cite{ volokitin2023optimized}
\begin{equation} \label{eq:asym_t_sync}
T_\mathrm{S}(\chi_e)
=\begin{cases}
\frac{5}{2\sqrt{3}}, ~ &\text{if $\chi_e\to 0$}\\
1.46\chi_e^{-1/3},~ &\text{if $\chi_e \gg 1$}.
\end{cases}
\end{equation}
We derive the low-energy limit in App.~\ref{app:classical_limit}.

\subsection{Breit-Wheeler process}

A photon of quantum parameter $\chi_\gamma$ produces pairs at the total rate
\begin{equation} \label{eq:rate_bw}
  \frac{dN_\mathrm{BW}}{dt}=\frac{\alpha_f m_e^2 c^4}{\hbar \varepsilon_\gamma} \chi_\gamma T_\mathrm{BW}(\chi_\gamma),
\end{equation}
where the auxiliary function is
\begin{equation} \label{eq:t_bw_def}
\begin{split}
    T_\mathrm{BW}(\chi_\gamma)=\frac{1}{\pi \sqrt{3} \chi^2_{\gamma} }\int_0^{\chi_\gamma}\bigg[&\int_{\xi} ^\infty K_{1/3}(s)ds\\&- \left(2-\frac{3\chi_\gamma \xi}{2}\right)K_{2/3}(\xi)\bigg]d\chi_e,
\end{split}
\end{equation}
with the same $\xi$-parameter as in Eq.~\eqref{eq:dn_bw}.
The asymptotic limits of the auxiliary function are \cite{ volokitin2023optimized}
\begin{equation} \label{eq:asym_t_bw}
T_\mathrm{BW}(\chi_\gamma)
=\begin{cases}
\frac{3}{16}\sqrt{\frac{3}{2}}\exp{\left(-\frac{8}{3\chi_\gamma}\right)}, ~ &\text{if $\chi_\gamma\to 0$}\\
0.38\chi_\gamma^{-1/3},~ &\text{if $\chi_\gamma \gg 1$}.
\end{cases}
\end{equation}
The coefficient of the low-energy limit is $\frac{3}{16}\sqrt{3/2}\approx0.2296$.

\section{Implementation} \label{sect:implementation}

Here we build the MC algorithm from the spectra of Sect.~\ref{sect:theory} and construct the approximations that make the algorithm run without lookup tables.

\subsection{Monte Carlo sampling}

Established MC sampling algorithms for the strong-field QED processes generate two quantities at random \cite{lobet2016modeling, niel2018quantum, kirk2014modelling, duclous2011monte}:
i) the time, or path length, that an incoming particle propagates in the field before the process occurs; and
ii) the $\chi$-parameters of the particles the process produces.

The propagation length follows from a random number $\eta \in (0,1)$, which fixes the final optical depth
\begin{equation}
  \tau_f = -\ln(\eta)
\end{equation}
that the particle must reach for the process to occur.
Starting from $\tau_0 =0$, the optical depth grows over a time step $dt$ as
\begin{equation}
  \tau_{n+1} = \tau_n + \left(\frac{d\tau_n}{dt}\right)dt,
\end{equation}
at the rate set by the total rate of the process, $d\tau  / dt = dN / dt$.
Equations~\eqref{eq:rate_sync} and \eqref{eq:rate_bw} show that this growth rate is controlled by the auxiliary functions $T_\mathrm{S}$ and $T_\mathrm{BW}$, which is why we approximate them in elementary functions below.
Once $\tau_n \geq \tau_f$, the process occurs and the algorithm stores the quantum parameter $\chi$ of the incoming particle.

The produced particles then need their own quantum parameters, which follow from a second random number $\zeta\in(0,1)$.
For an incoming particle of quantum parameter $\chi$ and a produced particle of quantum parameter $\chi_1$, the cumulative probability is%
\footnote{Some variants of the algorithm \cite{martinez2018radiative, fedeli2022picsar} define the cumulative probability of synchrotron radiation by replacing the production spectrum with the radiation power spectrum of Eq.~\eqref{eq:prad} in the integrals, which regulates the infrared divergence of the photon emission spectrum.}
\begin{equation} \label{eq:cumulative_p}
  p(\chi, \chi_1) = \frac{\int_0^{\chi_1} \frac{d^2N}{d\chi' dt}\,d\chi'} {\int_0^{\chi} \frac{d^2N}{d\chi' dt}\,d\chi'},
\end{equation}
where $\chi_1$ is the quantum parameter of the photon for synchrotron radiation and of the electron for the Breit-Wheeler process.
The upper bound $\chi_1$ at which the probability reaches the random number $\zeta$ becomes the quantum parameter of the produced particle,
\begin{equation} \label{eq:p_inverse}
  p(\chi, \chi_1) = \zeta~
                  \Leftrightarrow~ \chi_1
                  = p^{-1}(\chi, \zeta),
\end{equation}
and the second produced particle takes the remainder,
\begin{equation}
  \chi_2 = \chi-\chi_1.
\end{equation}
The cumulative probabilities are too involved to invert analytically, so simulations perform the inversion numerically or by tabulation.
We instead approximate $p(\chi, \chi_1)$ itself by a form that is both accurate and invertible in closed form, which yields an analytic formula for $\chi_1$.

Approximating the probability rather than the numerical solution for $\chi_1$ may look like a detour, but it has two advantages.
First, the probability connects directly to a physical quantity, the spectrum of the process, so the approximation can be tested against theory.
Differentiating Eq.~\eqref{eq:cumulative_p} and writing the total rate through the auxiliary function expresses the spectrum as the derivative of the probability,
\begin{equation} \label{eq:spectrum_mc}
  \frac{d^2N}{d\chi_1 dt} = \frac{\alpha_f m_e ^2 c^4T(\chi)}{\hbar\varepsilon} \frac{dp(\chi, r)}{dr},
\end{equation}
where we changed variables to $r=\chi_1/\chi$.
The accuracy of this spectrum measures the quality of the approximation.
For synchrotron radiation we compare against the radiation power spectrum instead of the emission spectrum;
using Eq.~\eqref{eq:prad},
\begin{equation} \label{eq:prad_mc}
  \frac{dP_\mathrm{rad}}{d\chi_\gamma} = \frac{\alpha_f m_e ^2 c^4rT_\mathrm{S}(\chi_e)}{\hbar} \frac{dp_\mathrm{S}(\chi_e, r)}{dr}.
\end{equation}
The auxiliary functions $T(\chi)$ in Eqs.~\eqref{eq:spectrum_mc} and \eqref{eq:prad_mc} are the numerically calculated ones, not their approximations.
Second, known physical properties of the spectra, such as asymptotic behavior or symmetries, can be imposed on the approximation directly.
Section~\ref{sect:results} shows that the $\chi_1$-distributions obtained by inverting the approximated probabilities coincide with the theoretical ones for a large number of monoenergetic incoming particles.

Tables~\ref{tab:t_params}, \ref{tab:sync_params}, and \ref{tab:bw_params} list the parameters of all approximations to five digits, which is sufficient for the accuracy the simulations require.
The supplementary implementation code%
\footnote{\url{https://github.com/hel-astro-lab/External-field-QED-Monte-Carlo-implementation}}
includes the same parameters to higher precision.

\subsection{Fitting by error minimization}

Each approximation $f(\chi)$ is a function with a number of free parameters fixed by optimizing a chosen property.
For a target function $y(\chi)$, we obtain the parameters of $f(\chi)$ by minimizing the least-squares-like error sum
\begin{equation} \label{eq:err_sum}
  \mathrm{Err}(y, f)=\sum_{i}\frac{(y(\chi_i)-f(\chi_i))^2}{\sigma_i}.
\end{equation}
The weights $\sigma_i = \sigma(\chi_i)$ can be used to control which regions dominate the fit.
Taking the reciprocal of the particle production spectrum as the weight, for instance, gives the regions of copious particle production the largest influence.

When comparing the pair production or radiation power spectra with their approximations, we drop the physical constants $\alpha_f m_e^2 c^4/\hbar$ and the energies $\varepsilon$ from the front and compare only the dimensionless parts;
the figures show these dimensionless quantities.
The dropped prefactor is 
\begin{equation}
  \frac{\alpha_f m_e^2 c^4}{\hbar \varepsilon}=\frac{\alpha_fm_ec^2}{\hbar \gamma}
  \approx 5.67 \times 10^{18}~\frac{1}{\gamma}\mathrm{s}^{-1},
\end{equation}
where, $\gamma$ is the normalized energy of the incoming particle. The prefactors therefore scale the rates to $\sim10^{18}~ \mathrm{s}^{-1}$ and damp them by the factor $1/\gamma$.

\subsection{Padé approximants}

A Padé approximant is formally a ratio of two power series expanded around a point $x_0$ \cite{iterativemethodspade},
\begin{align}
  A_m(x, x_0) &=\sum_{i=0}^ma_i(x-x_0)^i, \\
  B_n(x,x_0)  &=\sum_{i=0}^n b_i(x-x_0)^i,
\end{align}
and the $[m,n]$-order approximant is
\begin{equation}
  R_{n}^{m}(x, x_0)=\frac{A_m(x, x_0)}{B_n(x, x_0)}.
\end{equation}
In the formal definition, the coefficients $a_i$ and $b_i$ are fixed so that the approximant reproduces the first $m+n$ terms of the Taylor series
\begin{equation}
  R_n^m(x, x_0)=\sum_{i=0}^{m+n} c_i (x-x_0)^i.
\end{equation}
We use the term less formally for any rational function of polynomials.
Such approximants are flexible and remain analytically invertible as long as $m,n\leq 4$, which makes them the natural choice for the cumulative probabilities.
We write the probabilities in terms of $r=\chi_1/\chi$ as approximants expanded around $r=1$,
\begin{equation}
  p(\chi,r)\approx1- R_n^m(-r,-1).
\end{equation}

\subsection{Auxiliary function of synchrotron radiation}

\begin{figure}[t]
    \centering
    \includegraphics[scale=0.73, clip=true, trim={0.35cm 1.4cm 0.2cm 1.05cm}]{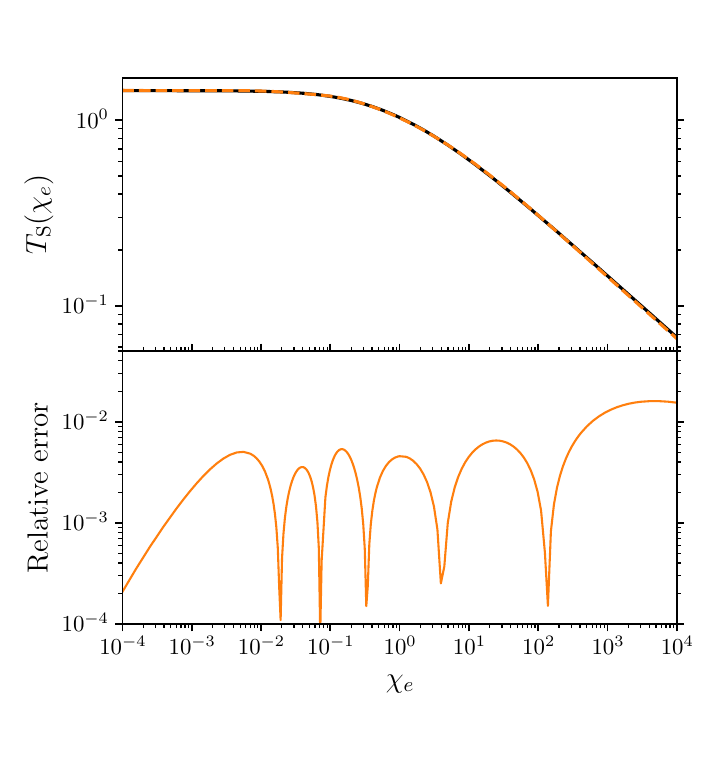}
    \caption{Auxiliary function of synchrotron radiation and its approximation. Top: $T_\mathrm{S}(\chi_e)$ from numerical integration of Eq.~\eqref{eq:t_sync_def} (solid black) and the approximation of Eq.~\eqref{eq:t_sync_approx} (dashed orange). Bottom: relative error of the approximation. 
    }\label{fig:aux_t_sync}
\end{figure}

We approximate the double Bessel integral in the expression of $T_\mathrm{S}$ with a product of three elementary functions.
To reproduce the asymptotic limits of Eq.~\eqref{eq:asym_t_sync}, we define
\begin{equation} \label{eq:t_sync_approx}
  T^{\mathrm{Appr}}_\mathrm{S}(\chi_e)=\frac{5}{2\sqrt{3}}f(\chi_e) C_1(\chi_e) C_2(\chi_e),
\end{equation}
where $f(\chi_e)$ carries the limits
\begin{equation}
f(\chi_e)\to
\begin{cases}
    1,~ &\chi_e\to 0\\
    \frac{2\sqrt{3}}{5} \times 1.46\chi_e^{-1/3},~ &\chi_e \to \infty.
\end{cases}
\end{equation}
One function with these limits is
\begin{equation}
  f(\chi_e)=1-\exp{\left(-\frac{2\sqrt{3}}{5}\cdot 1.46\chi_e^{-1/3}\right)},
\end{equation}
and $C_1(\chi_e)$ and $C_2(\chi_e)$ are correction functions with three free parameters each.
Minimizing the error sum of Eq.~\eqref{eq:err_sum} gives the correction functions
\begin{align}
  C_1(\chi_e) &=1 + \alpha_0\chi_e \exp{\left(-\frac{\left(\chi_e+\delta_0\right)^2}{\Delta_0}\right)}, \label{eq:sync_c1} \\
  C_2(\chi_e) &=1 + \alpha_1\chi_e \exp{\left(-\frac{\left(\ln(\chi_e)+\delta_1\right)^2}{\Delta_1}\right)}, \label{eq:sync_c2}
\end{align}
whose parameter values are listed in Table~\ref{tab:t_params}.
Figure~\ref{fig:aux_t_sync} compares the approximation with the exact function:
the relative error stays below $0.7\%$ across the fitted range $\chi_e\in(10^{-3},10^{2})$, peaking at $\chi_e\approx25$, and grows to $1.6\%$ near $\chi_e = 5\times10^3$, far above the range where synchrotron emission usually dominates in simulations.

\subsection{Cumulative probability of synchrotron radiation}

\begin{figure}[t]
    \centering
    \includegraphics[scale=0.59, clip=true, trim={0.55cm 2.5cm 0.25cm 0}]{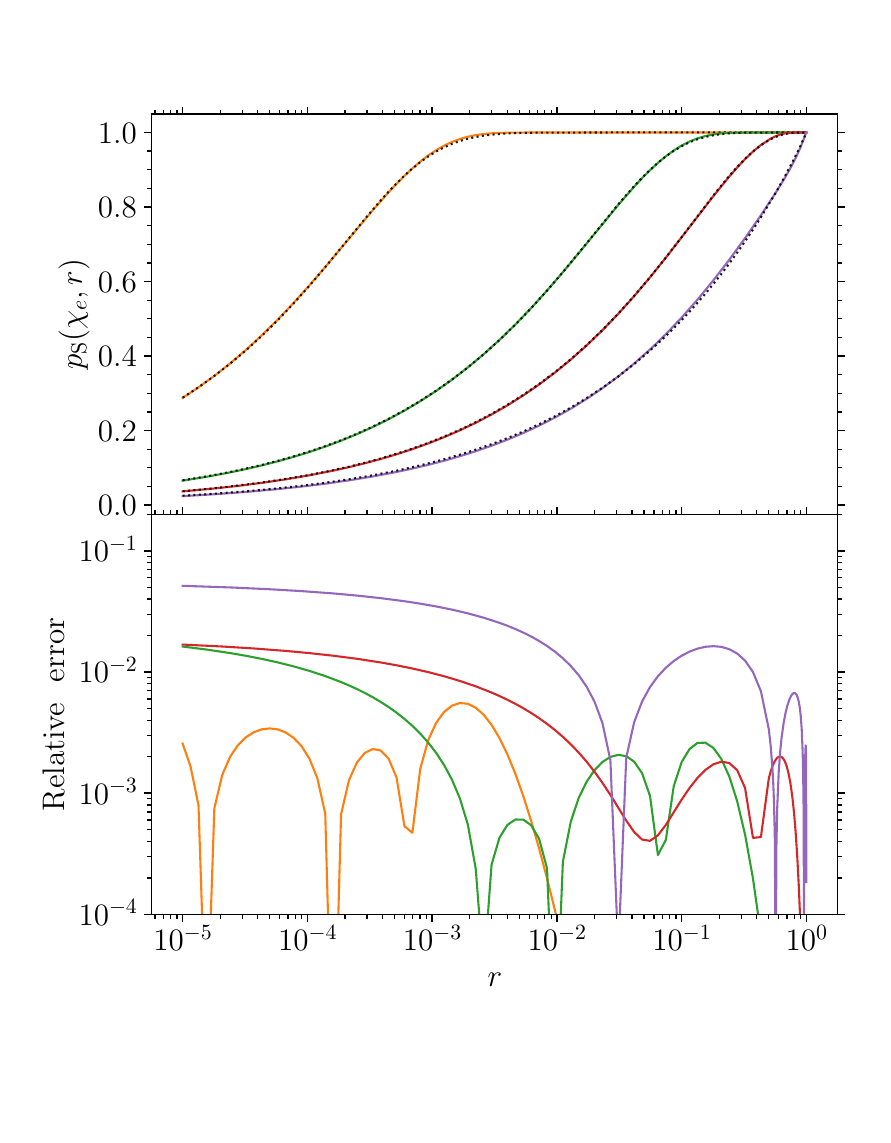}
    \caption{Cumulative probability of synchrotron radiation and its approximation. Top: $p_\mathrm{S}(\chi_e,r)$ against the energy fraction $r=\chi_\gamma/\chi_e$, computed numerically from Eq.~\eqref{eq:cumulative_p} (solid curves) and from the approximation of Eq.~\eqref{eq:sync_pade} (dotted curves), for $\chi_e = 0.0005$ (orange), $0.05$ (green), $0.5$ (red), and $50$ (purple). Bottom: relative error for each $\chi_e$.}
    \label{fig:sync_prob}
\end{figure}

\begin{figure}[t]
    \centering
    \includegraphics[scale=0.59, clip=true, trim={0.55cm 2.5cm 0.25cm 1.15cm}]{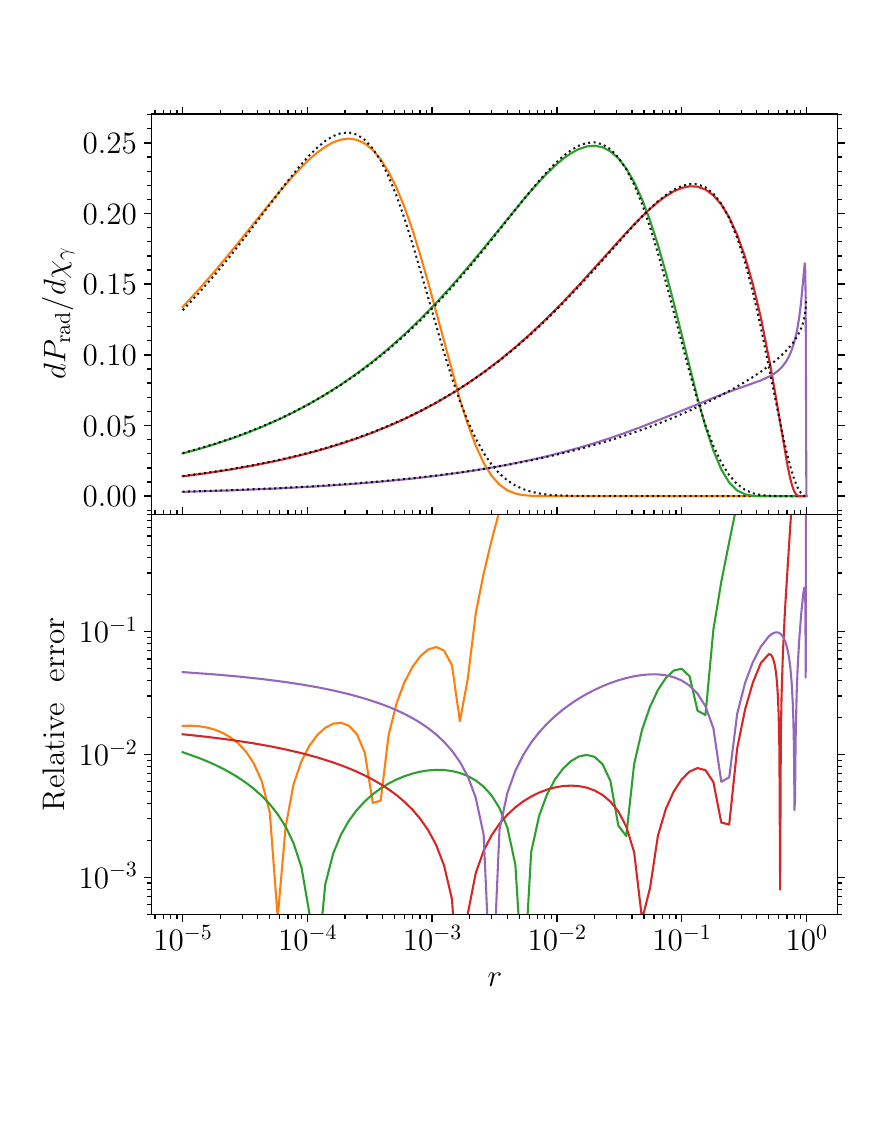}
    \caption{Radiation power spectrum recovered from the approximated probability. Top: $dP_\mathrm{rad}/d\chi_\gamma$ computed numerically from Eq.~\eqref{eq:prad} (solid curves) and from the derivative of the approximation through Eq.~\eqref{eq:prad_mc} (dotted curves), for the same four $\chi_e$ values as in Fig.~\ref{fig:sync_prob}. Bottom: relative error. For $\chi_e<1$ the error exceeds a percent only where the spectrum itself is negligibly small. In the $\chi_e >1$ regime the derivatives peak slower than the numerical values around $r\to 1$, however, the difference has negligible effects on the approximation of the cumulative probability which can be seen in Fig. \ref{fig:sync_prob}.
    }
    \label{fig:sync_spectrum}
\end{figure}

\begin{figure}[ht!]
    \centering
    \includegraphics[scale=0.6, clip=true, trim={0.75cm 4.05cm 0.0cm 1.5cm}]{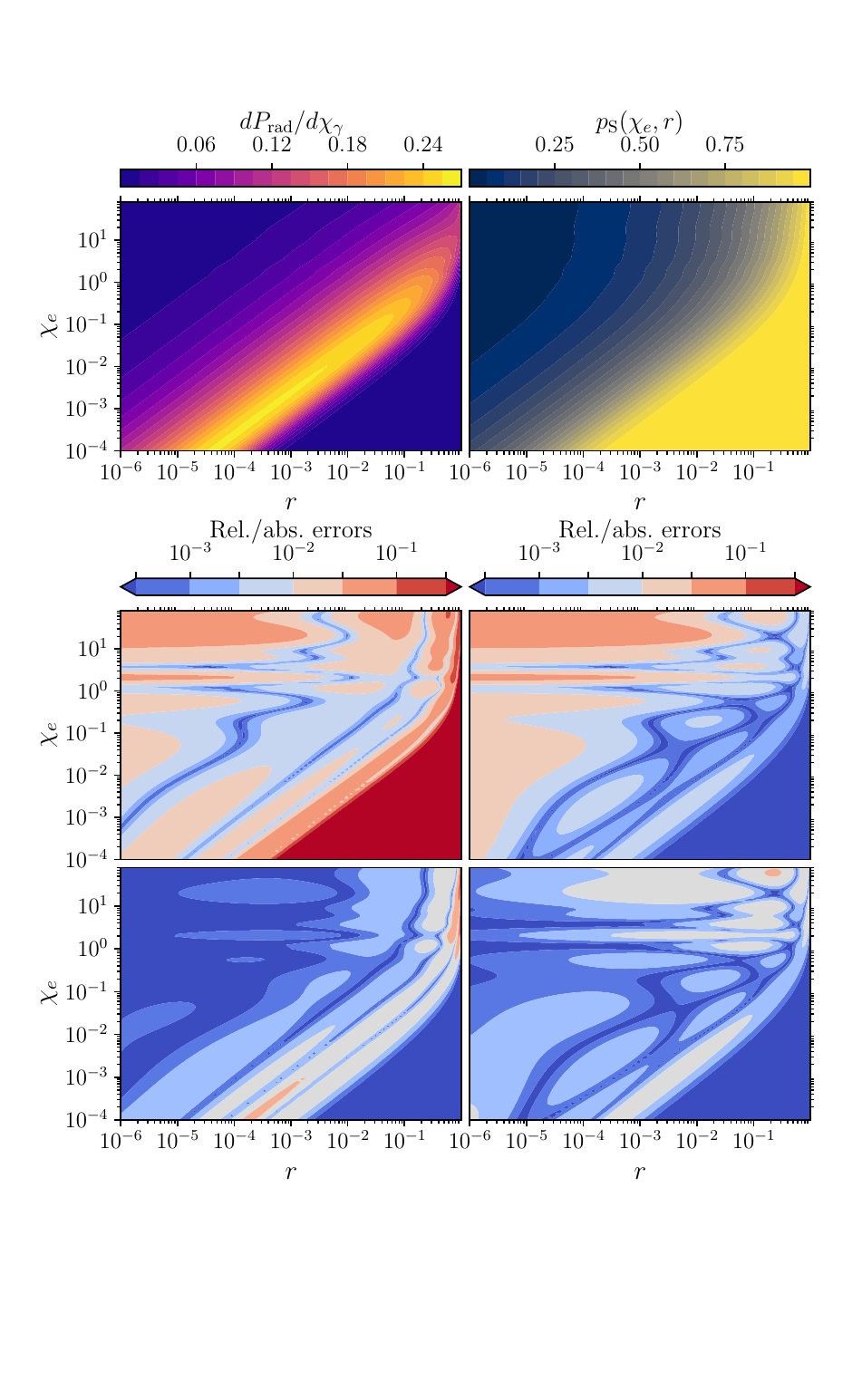}
    \caption{
    Accuracy of the synchrotron approximations over the $(r,\chi_e)$ plane. 
    Left: radiation power spectrum $dP_\mathrm{rad}/d\chi_\gamma$; 
    right: cumulative probability $p_\mathrm{S}(\chi_e,r)$. 
    Top row: values given by the approximations; middle and bottom rows: relative and absolute error against the numerical values. Both functions are vanishingly small in most areas where the relative error is large, so the absolute error stays small there.}
    \label{fig:sync_2d}
\end{figure}

Next we build an approximant for the cumulative probability of the synchrotron emission.
To reach sufficient accuracy while retaining invertibility, we approximate the cumulative probability of synchrotron radiation with a fourth-order Padé approximant in the ratio $r=\chi_\gamma / \chi_e$.
With the shorthand $x= (1-r^{1/3})^n$, our approximation reads
\begin{equation} \label{eq:sync_pade}
  p_\mathrm{S}(\chi_e, r)=1 -\frac{ax^4}{1 + bx^2 + cx^3}.
\end{equation}
The power $1/3$ in $r$ enforces the correct power-law behavior at small $\chi_\gamma$ \cite{Hofmann:202177}, and the auxiliary parameters $a$, $b$, $c$, and $n$ are functions of $\chi_e$.
The form satisfies $p_\mathrm{S}(\chi_e, 1)=1$ by construction, and demanding $p_\mathrm{S}(\chi_e,0)=0$,
fixes the parameter $c=a-b-1$.
The derivative
\begin{equation} \label{eq:sync_p_deriv}
  r\frac{dp_\mathrm{S}}{dr} = \frac{anr^{1/3}}{3}\frac{(1-r^{1/3})^{4n-1}\left(4 + 2bx^2 + cx^3\right)}{\left(1 + bx^2 + cx^3\right)^2}
\end{equation}
gives the radiation power spectrum through Eq.~\eqref{eq:prad_mc}, which we compare with the numerical values from Eq.~\eqref{eq:prad}.
The radiation spectrum vanishes as $r\to 1$ provided that $n(\chi_e) > 1/4.$ App.~\ref{app:param_sync} describes how we determined the auxiliary parameters and their approximating functions. Figures~\ref{fig:sync_prob} and \ref{fig:sync_spectrum} compare the approximated probability and the resulting radiation power spectrum with the numerical values, and Fig.~\ref{fig:sync_2d} maps the errors over the whole $(r,\chi_e)$ plane.

Sampling the photon quantum parameter means equating the approximated cumulative probability to the random number $\zeta \in (0,1)$ as $\zeta=p_\mathrm{S}(\chi_e,r)$,
which reads
\begin{equation} \label{eq:sync_randomzeta}
  \zeta=1-\frac{ax^{4}}{1+bx^{2}+cx^{3}}.
\end{equation}
Inverting the probability as in Eq.~\eqref{eq:p_inverse} amounts to solving this equation;
the details of the solution are given in App.~\ref{App:QuarticSyncEq}.
The resulting quantum parameter of the emitted photon is,

\begin{equation} \label{eq:sync_chi_gamma}
  \chi_\gamma = \chi_e\left[1-\left(u-\frac{C}{4a}\right)^{1/n}\right]^3.
\end{equation}
where $u$ and $C$ are complicated expressions dependent on the auxiliary parameters, given in the Appendix \ref{App:QuarticSyncEq}.
A simulation evaluates Eq.~\eqref{eq:sync_chi_gamma} directly for a given $\chi_e$ and $\zeta$, and the conservation law fixes the outgoing electron,
$ \chi_{e'}=\chi_e - \chi_\gamma$.

\begin{figure}[t]
    \centering
    \includegraphics[scale=0.65, clip=true, trim={0.3cm 2.35cm 0.5cm 0.9cm}]{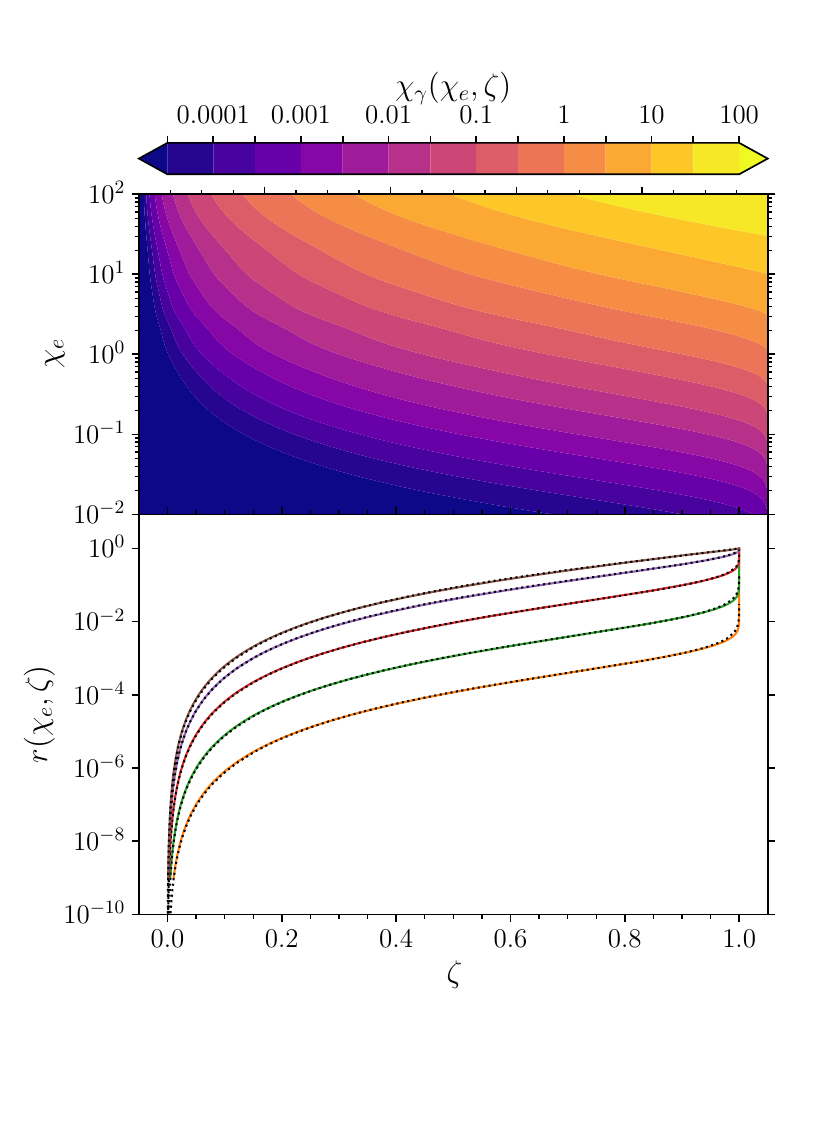}
    \caption{Photon quantum parameters sampled from the inverted probability. 
    Top: $\chi_\gamma(\chi_e,\zeta)$ from Eq.~\eqref{eq:sync_chi_gamma} over the plane of random number $\zeta$ and electron quantum parameter $\chi_e$. 
    Bottom: energy fraction $r=\chi_\gamma/\chi_e$ from Eq.~\eqref{eq:sync_r} (dotted curves) against the numerically inverted values (solid curves) for $\chi_e = 0.001$ (orange), $0.01$ (green), $0.1$ (red), $1$ (purple), and $10$ (brown).}
    \label{fig:sync_chi_sol}
\end{figure}

Figure~\ref{fig:sync_chi_sol} shows the sampled photon quantum parameter over the $(\zeta,\chi_e)$ plane together with the energy fractions it produces.
At small $\chi_e$ the fractions stay well below unity even for $\zeta\to1$, whereas the numerical values continue to rise.
This behavior reflects the classical limit:
emitted photons carry a vanishing probability of taking a significant share of the electron energy, and the radiation becomes continuous.

\subsection{Auxiliary function of the Breit-Wheeler process}

\begin{figure}[ht]
    \centering
    \includegraphics[scale=0.8, clip=true, trim={0.5cm 1.75cm 0cm 0.5cm}]{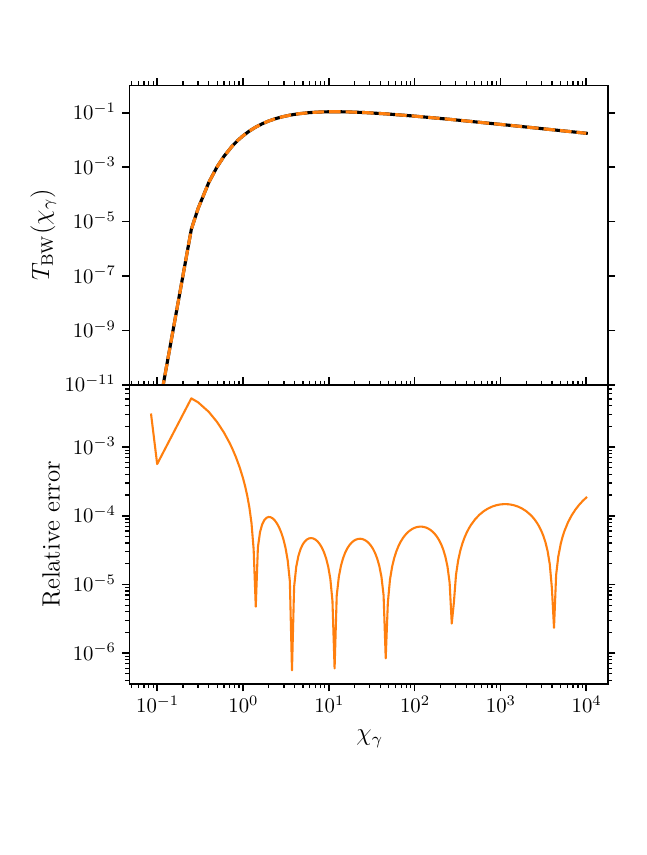}
    \caption{Auxiliary function of the Breit-Wheeler process and its approximation. Top: $T_\mathrm{BW}(\chi_\gamma)$ from numerical integration of Eq.~\eqref{eq:t_bw_def} (solid black) and the approximation of Eq.~\eqref{eq:t_bw_approx} (dashed orange). Bottom: relative error, obtained with unit weights $\sigma_i = 1$ in the error sum.}
    \label{fig:aux_t_bw}
\end{figure}

We approximate the double Bessel integral in the expression of $T_\mathrm{BW}$ following a same technique as for the $T_\mathrm{S}$.
To reproduce the asymptotic limits of Eq.~\eqref{eq:asym_t_bw}, we write the approximation of $T_\mathrm{BW}$ as
\begin{equation}
  T^\mathrm{Appr}_\mathrm{BW}(\chi_\gamma)=0.38 \chi_\gamma^{-1/3}\exp{\left(-\frac{8}{3\chi_\gamma}\right)}f(\chi_\gamma) C(\chi_\gamma),
\end{equation}
where $f(\chi_\gamma)$ carries the limits
\begin{equation}
f(\chi_\gamma) \approx
\begin{cases}
    1,~ &\chi_\gamma\to \infty\\
    A\chi_\gamma^{1/3},~ &\chi_\gamma \to 0
\end{cases}
\end{equation}
and $A=\frac{3}{16}\sqrt{3/2}/0.38\approx0.60$, and $C(\chi_\gamma)$ is a correction function with three free parameters.
Minimizing the error sum of Eq.~\eqref{eq:err_sum} gives
\begin{equation} \label{eq:t_bw_approx}
  T_\mathrm{BW}^{\mathrm{Appr}}(\chi_\gamma)=0.3778 \chi_\gamma^{-0.333}\exp{\left(-\frac{8}{3\chi_\gamma}\right)}f(\chi_\gamma)C(\chi_\gamma),
\end{equation}
with
\begin{equation} \label{eq:bw_f}
  f(\chi_\gamma) = 1-\exp{\left(-\Delta_6\chi_\gamma^{k_9}\right)}
\end{equation}
and the correction function
\begin{equation} \label{eq:bw_c}
  C(\chi_\gamma)=1-\alpha_6\exp{\left(-\frac{(\delta_5+\ln{(\chi_\gamma)})^2}{\Delta_7}\right)}.
\end{equation}
Table~\ref{tab:t_params} lists the parameters of $f(\chi_\gamma)$ and $C(\chi_\gamma)$.
As Fig.~\ref{fig:aux_t_bw} shows, this approximation is the most accurate of the four:
the relative error stays below $3.3\times10^{-4}$ for $\chi_\gamma \gtrsim 1$ and reaches $5.5\times10^{-3}$ only near $\chi_\gamma \approx 0.2$, where the pair-production rate is exponentially suppressed.
The fitted exponent $k_9$ and amplitude $\Delta_6$ do not reproduce the $\chi_\gamma\to0$ behaviour that the functional form was built to carry:
they give $T^\mathrm{Appr}_\mathrm{BW}\to0.76\chi_\gamma^{-0.064}e^{-8/3\chi_\gamma}$ instead of $\frac{3}{16}\sqrt{3/2}\,e^{-8/3\chi_\gamma}$. We use such normalization value that does not reproduce the small-parameter limit because it results in a smaller overall error when the correction function Eq.~\eqref{eq:bw_c} is used.

\begin{table}[ht]
\centering
\begin{tabular}{llllll}
     & $T_\mathrm{S}^\mathrm{Appr}(\chi_e)$ &  &  & $T_\mathrm{BW}^\mathrm{Appr}(\chi_\gamma)$ & \\ \hline
     Eq. & Var. & Value & Eq. & Var. & Value \\ \hline
     & $\alpha_0$ & -403289.42146 & \eqref{eq:bw_f} & $\Delta_6$ & 2.00326 \\
     \eqref{eq:sync_c1} & $\delta_0$ & 0.84734 &  & $k_9$ & 0.26880 \\ \cline{4-6}
     & $\Delta_0$ & 0.05843 &  & $\alpha_6$ & 0.63309 \\ \cline{1-3}
     & $\alpha_1$ & 3.26342 & \eqref{eq:bw_c} & $\delta_5$ & 3.93232 \\
     \eqref{eq:sync_c2} & $\delta_1$ & 7.95554 &  & $\Delta_7$ & 29.94554 \\
     & $\Delta_1$ & 19.18831 &  &  &
\end{tabular}
\caption{Parameters of the approximated auxiliary functions $T_\mathrm{S}^\mathrm{Appr}$ and $T_\mathrm{BW}^\mathrm{Appr}$.}
\label{tab:t_params}
\end{table}

\subsection{Cumulative probability of the Breit-Wheeler process}

\begin{figure}[ht]
    \centering
    \includegraphics[scale=0.57, clip=true, trim={0.25cm 2.25cm 0.25cm 1.5cm}]{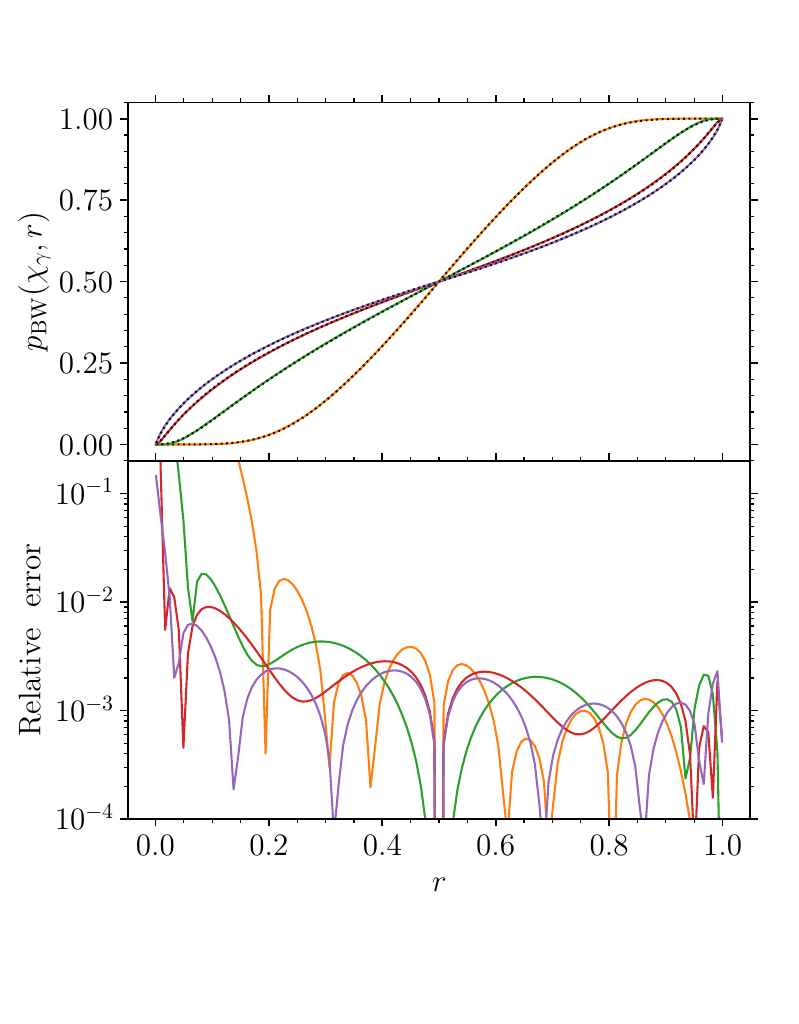}
    \caption{Cumulative probability of the Breit-Wheeler process and its approximation. Top: $p_\mathrm{BW}(\chi_\gamma,r)$ against the energy fraction $r=\chi_e/\chi_\gamma$, computed numerically from Eq.~\eqref{eq:cumulative_p} (solid curves) and from the approximation of Eq.~\eqref{eq:bw_pade} (dotted curves), for $\chi_\gamma=1$ (orange), $10$ (green), $100$ (red), and $1500$ (purple). Bottom: relative error. The large errors as $r\to0$ affect vanishingly small probabilities and never enter the sampling, which uses Eq.~\eqref{eq:bw_symmetry} below $r=0.5$.}
    \label{fig:bw_prob}
\end{figure}

\begin{figure}[ht]
    \centering
    \includegraphics[scale=0.57, clip=true, trim={0.25cm 2.25cm 0.25cm 1.5cm}]{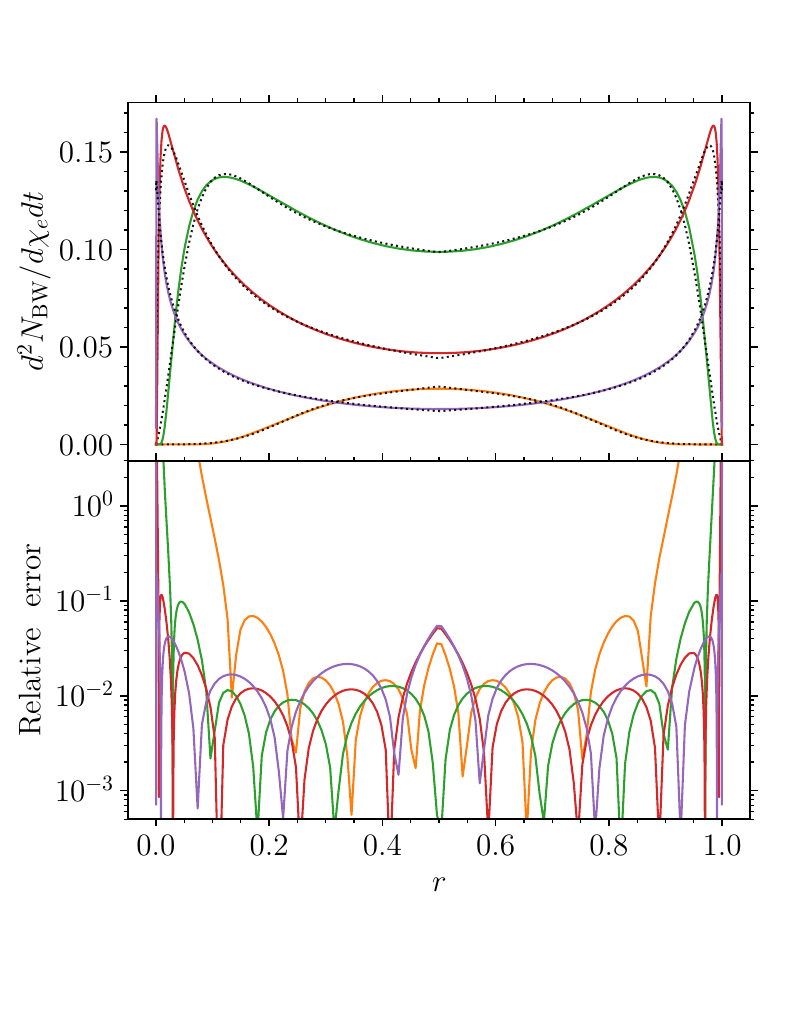}
    \caption{Pair-production spectrum recovered from the approximated probability. Top: $d^2N_\mathrm{BW}/d\chi_e dt$ computed numerically (solid curves) and from the derivative of the approximation through Eq.~\eqref{eq:spectrum_mc} (dotted curves), for the same four $\chi_\gamma$ values as in Fig. \ref{fig:bw_prob}. Bottom: relative error. Values below $r=0.5$ follow from manual symmetrization, Eq.~\eqref{eq:bw_symmetry}.}
    \label{fig:bw_spectrum}
\end{figure}

\begin{figure}[ht]
    \centering
    \includegraphics[scale=0.69, clip=true, trim={0.5cm 3.25cm 0.75cm 1.5cm}]{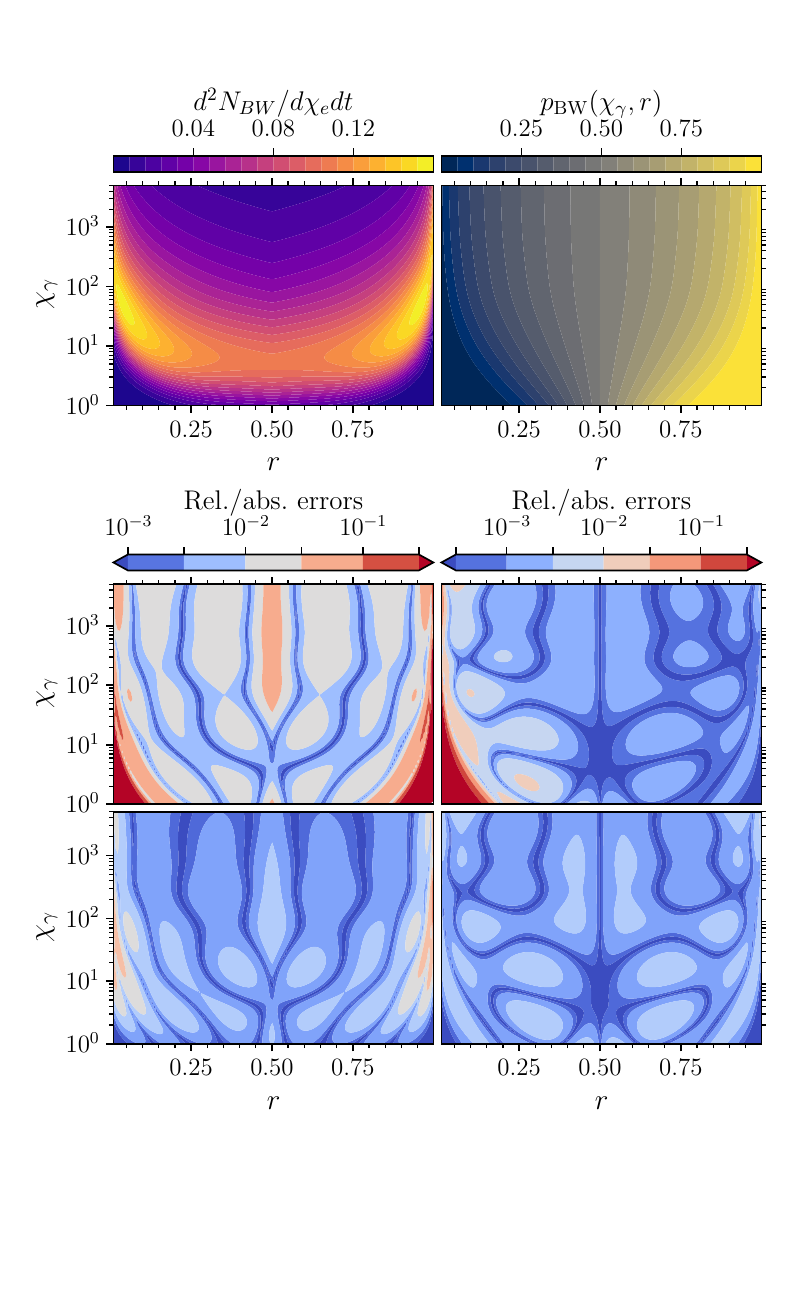}
    \caption{Accuracy of the Breit-Wheeler approximations over the $(r,\chi_\gamma)$ plane. Left column: pair-production spectrum $d^2N_\mathrm{BW}/d\chi_e dt$; right column: cumulative probability $p_\mathrm{BW}(\chi_\gamma,r)$. Top row: values given by the approximations; middle and bottom rows: relative and absolute error against the numerical values. Values below $r=0.5$ follow from manual symmetrization, Eq.~\eqref{eq:bw_symmetry}.}
    \label{fig:bw_2d}
\end{figure}

Next we build an approximant for the cumulative probability of the Breit-Wheeler process.
The Breit-Wheeler spectrum is symmetric about $r=\chi_e/\chi_\gamma=1/2$, as Figs.~\ref{fig:bw_prob} and \ref{fig:bw_spectrum} show, because the electron and the positron are interchangeable.
The approximation therefore only needs to cover $r\in [0.5,1)$;
the values at $r<0.5$ follow from the symmetry
\begin{equation} \label{eq:bw_symmetry}
  p_\mathrm{BW}(\chi_\gamma,r)=1-p_\mathrm{BW}(\chi_\gamma,1-r).
\end{equation}
We call this manual symmetrization:
the approximation itself need not be symmetric, provided that it passes through the point $(0.5, 0.5)$ and stays accurate on the upper half of the interval.
As for synchrotron radiation, we use a fourth-order Padé approximant, here in $r=\chi_e/\chi_\gamma$ and with the shorthand $u=(1-r)^n$,
\begin{equation} \label{eq:bw_pade}
  p_\mathrm{BW}(\chi_\gamma,r)=1-\frac{au^4}{1+bu+cu^2}.
\end{equation}
Demanding $p_\mathrm{BW}(\chi_\gamma,1/2)=1/2$ fixes
\begin{equation}
  c = \frac{2a- 2^{3n}b-2^{4n}}{2^{2n}},
\end{equation}
which is also what makes Eq.~\eqref{eq:bw_symmetry} usable.
We determine the free auxiliary parameters $a(\chi_\gamma)$, $b(\chi_\gamma)$, and $n(\chi_\gamma)$ by comparing Eq.~\eqref{eq:bw_pade} with the numerical cumulative probability and its derivative
\begin{equation} \label{eq:bw_p_deriv}
  \frac{dp_\mathrm{BW}}{dr}=\frac{na(4+3bu+2cu^2)(1-r)^{4n-1}}{(1+bu+cu^2)^2}
\end{equation}
with the numerical pair-production spectrum through Eq.~\eqref{eq:spectrum_mc}.
The spectrum vanishes at the boundary $r\to 1$ only if $n(\chi_\gamma)>1/4$. App.~\ref{app:param_bw} describes how we determined the auxiliary parameters and their approximating functions. Figures~\ref{fig:bw_prob}--\ref{fig:bw_2d} compare the approximation with the numerical probability and spectrum.

\begin{figure}[h!]
    \centering
    \includegraphics[scale=0.6, clip=true, trim = {0.3cm 2.35cm 0.3cm 0}]{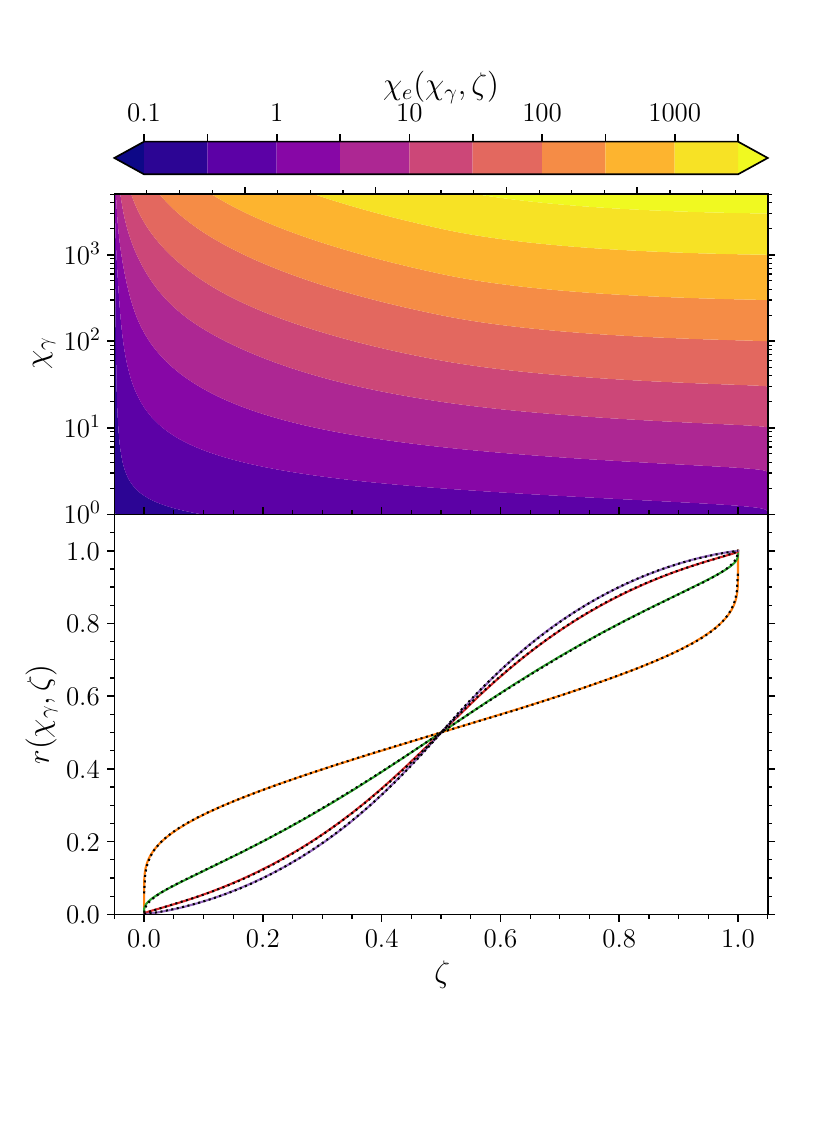}
    \caption{Electron quantum parameters sampled from the inverted probability. Top: $\chi_e(\chi_\gamma,\zeta)$ from Eq.~\eqref{eq:bw_chi_e} over the plane of random number $\zeta$ and photon quantum parameter $\chi_\gamma$. Bottom: energy fraction $r=\chi_e/\chi_\gamma$ from Eq.~\eqref{eq:bw_r} (dotted curves) against the numerically inverted values (solid curves) for $\chi_\gamma = 1$ (orange), $10$ (green), $100$ (red), and $1500$ (purple). Values at $\zeta<0.5$ follow from manual symmetrization.}
    \label{fig:bw_chi_sol}
\end{figure}

Sampling the electron quantum parameter again means equating the cumulative probability to the random number $\zeta \in (0,1)$ as $\zeta=p_\mathrm{BW}(\chi_\gamma,r)$,
where values $\zeta <0.5$ call for manual symmetrization as in Eq.~\eqref{eq:bw_symmetry}.
Written out, this is the quartic equation
\begin{equation} \label{eq:bw_randomzeta}
\zeta =   1-\frac{au^4}{1+bu+cu^2},
\end{equation}
Inverting the probability as in Eq.~\eqref{eq:p_inverse} amounts to solving this equation;
the detailed solution is given in App.~\ref{app:QuarticBWEq}. Combined with manual symmetrization, the resulting quantum parameter of the produced electron is
\begin{equation} \label{eq:bw_chi_e}
    \chi_e(\chi_\gamma, \zeta)=
    \begin{cases}
        \left(1-[u_2^+(\chi_\gamma, \zeta)]^{1/n}\right)\chi_\gamma,~&\text{for} ~\zeta\geq0.5\\
        [u_2^+(\chi_\gamma, 1-\zeta)]^{1/n}\chi_\gamma,~&\text{for} ~\zeta<0.5,
    \end{cases}
\end{equation}
where $u_2^+$ is a complicated expression given in terms of the auxiliary parameters in App. \ref{app:QuarticBWEq}. A simulation evaluates Eq. \eqref{eq:bw_chi_e} directly for a given $\chi_\gamma$ and $\zeta$, and the conservation law fixes the positron, $\chi_{e^+}=\chi_\gamma-\chi_e$.
Figure~\ref{fig:bw_chi_sol} shows the sampled electron quantum parameter and the energy fractions it produces.

\section{Results} \label{sect:results}

Two independent tests measure how well the approximations reproduce the physics.
The first compares the derivative of the approximated cumulative probability with the exact spectrum through Eq.~\eqref{eq:spectrum_mc}, which predicts the spectrum a simulation would produce without running one.
Figures~\ref{fig:sync_spectrum}, \ref{fig:sync_2d}, \ref{fig:bw_spectrum}, and \ref{fig:bw_2d} collect the outcome:
the predicted spectra deviate from the exact ones by about a percent, and the regions of larger relative error are typically those where the spectra are vanishingly small, which keeps the absolute error small there as well.
This test also guided the choice among candidate approximations during their development.
The second test, described below, samples the inverted approximations directly.

We draw $N\gg1$ random numbers $\zeta$ from a uniform distribution.
For synchrotron radiation, we evaluate $r(\chi_e, \zeta)$ from Eq.~\eqref{eq:sync_r} for each $\zeta$ at fixed $\chi_e$ and bin the results into a normalized histogram with $\sqrt{N}$ bins, which gives the distribution of photon quantum parameters that the approximation produces. The histogram is compared to the normalized spectra that Eq.~\eqref{eq:dn_s} predicts.
We repeat this for four values of $\chi_e$.
The Breit-Wheeler case follows the same procedure with $r(\chi_\gamma, \zeta)$ from Eq.~\eqref{eq:bw_r} at fixed $\chi_\gamma$, which gives the distribution of the quantum parameters of the produced electrons and positrons, again for four values of $\chi_\gamma$. The result is compared to the normalized spectra predicted by Eq. \eqref{eq:dn_bw}.

\begin{figure}[t!]
    \centering
    \includegraphics[scale=0.57, clip=true, trim = {1.25cm 1.5cm 0.5cm 0}]{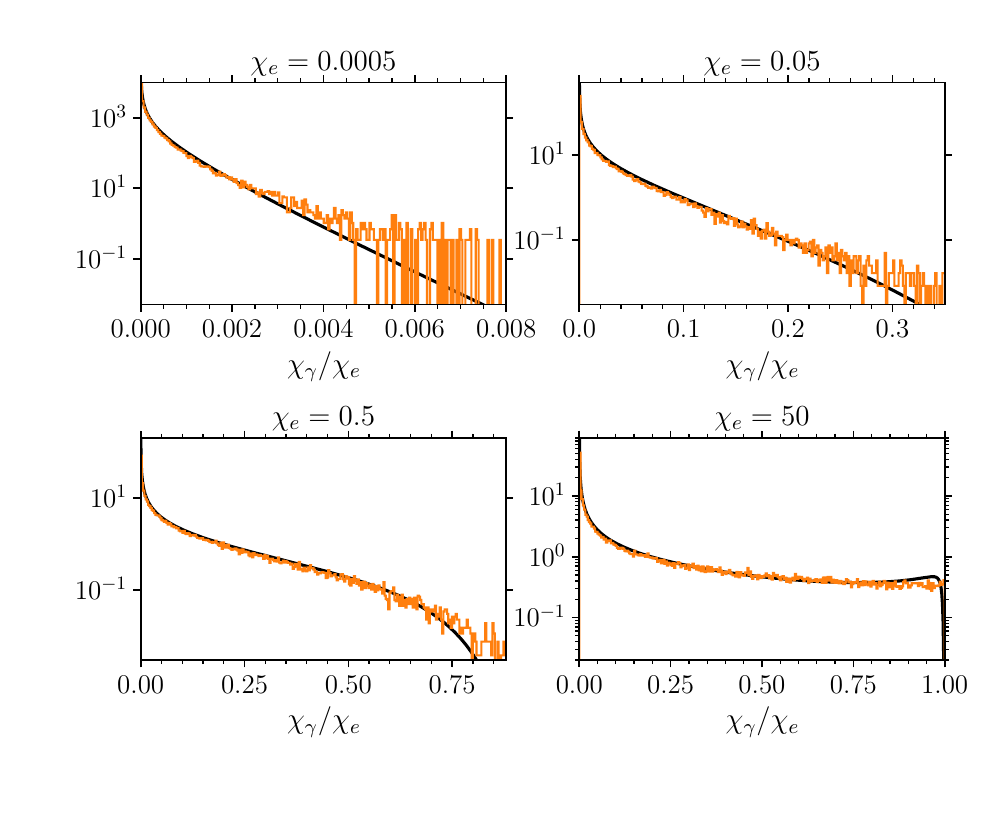}
    \caption{Distribution of photon quantum parameters sampled for synchrotron radiation. Normalized histograms of the $\chi_\gamma$ values emitted in $N=9\times10^{4}$ events by monoenergetic incoming electrons, sampled with Eq.~\eqref{eq:sync_r} (orange), against the density predicted by Eq.~\eqref{eq:dn_s} (solid black), for four values of $\chi_e$.}
    \label{fig:sync_sampled}
\end{figure}

\begin{figure}[t!]
    \centering
    \includegraphics[scale=0.57, clip=true, trim = {1.95cm 1.25cm 0 0}]{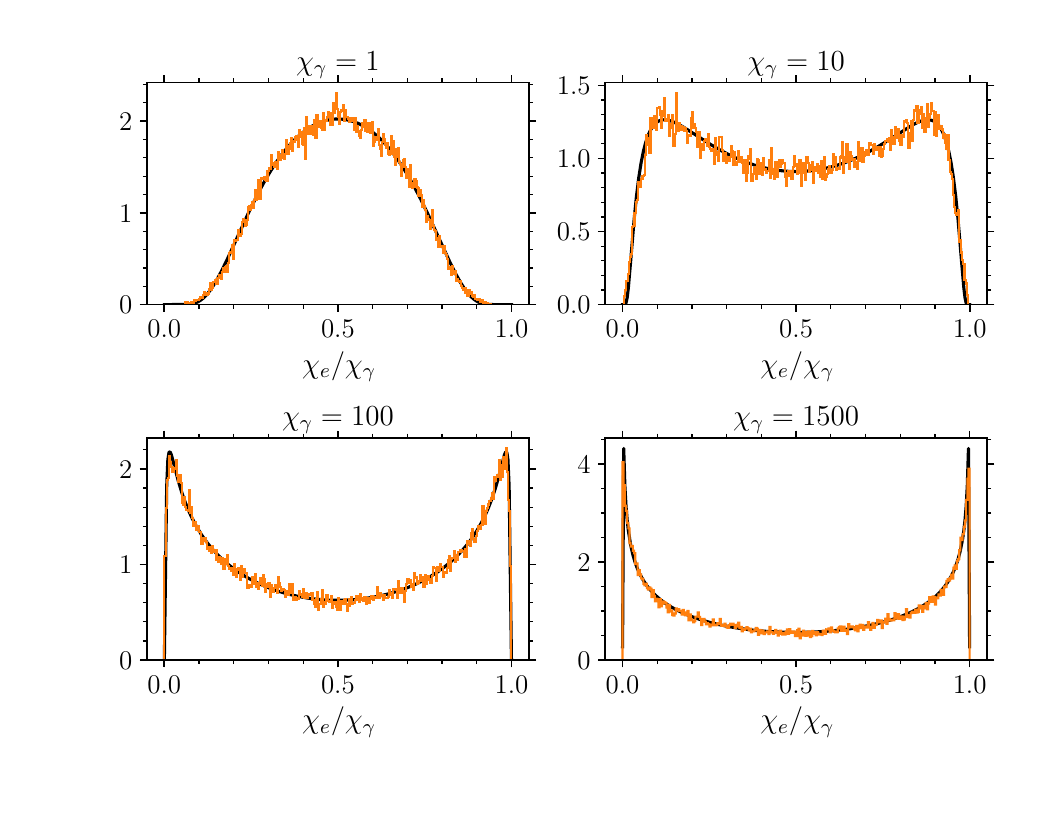}
    \caption{Distribution of electron quantum parameters sampled for the Breit-Wheeler process. Normalized histograms of the $\chi_e$ values produced in $N=9\times10^{4}$ events by monoenergetic incoming photons, sampled with Eq.~\eqref{eq:bw_r} (orange), against the density predicted by Eq.~\eqref{eq:dn_bw} (solid black), for four values of $\chi_\gamma$. Counts at $\zeta <0.5$ follow from manual symmetrization.}
    \label{fig:bw_sampled}
\end{figure}

Figures~\ref{fig:sync_sampled} and \ref{fig:bw_sampled} show that the sampled particle spectra follow the theoretical predictions for both processes.
For synchrotron radiation, part of the sampled photons carry energies orders of magnitude below anything Eq.~\eqref{eq:dn_s} resolves at $N=9\times10^{4}$.
This gap comes from the sample size rather than from a disagreement between approximation and theory:
the subintervals $(r_i, r_{i+1})$ must be uniform in size, so covering more decades in $r$ requires larger $N$.

\section{Discussion}\label{sect:discussion}

We have constructed approximations for the functions that Monte Carlo simulations of synchrotron radiation and of the nonlinear Breit-Wheeler process need:
the auxiliary functions $T_\mathrm{S}$ and $T_\mathrm{BW}$, which set when a process occurs, and the cumulative probabilities $p_\mathrm{S}$ and $p_\mathrm{BW}$, which set the energies of the particles it produces.
Inverting the approximated probabilities gives closed-form expressions for the quantum parameters of the produced particles, $\chi_\gamma$ and $\chi_e$. Every expression is built from elementary functions, so a simulation using them needs no lookup table, no interpolation between tabulated values, and no iterative inversion.

The approximation of $T_\mathrm{BW}$ agrees with the exact function to better than $3.3\times10^{-4}$ in relative terms for $\chi_\gamma \gtrsim 1$, and to $2.3\times10^{-3}$ across the whole fitted range $\chi_\gamma\in(0.5,5\times10^{3})$, degrading to $5.5\times10^{-3}$ near $\chi_\gamma \approx 0.2$.
The approximation of $T_\mathrm{S}$ is accurate to better than $0.7\%$ over the fitted range $\chi_e\in(10^{-3},10^{2})$ and to $1.6\%$ out to $\chi_e = 10^{4}$.

We tested the Padé approximants for $p_\mathrm{S}$ and $p_\mathrm{BW}$ in two ways.
Relating the derivative of the cumulative probability to the spectrum through Eq.~\eqref{eq:spectrum_mc} predicts errors of about a percent in the sampled spectra, as Figs.~\ref{fig:sync_spectrum}, \ref{fig:sync_2d}, \ref{fig:bw_spectrum}, and \ref{fig:bw_2d} show.
Sampling $N=9\times10^{4}$ events from the inverted approximations, Eqs.~\eqref{eq:sync_r} and \eqref{eq:bw_r}, reproduces the theoretical distributions in Sect.~\ref{sect:results}.
Both tests probe a single process in isolation, and neither follows how the errors accumulate over a full simulation.
A population of electrons that emits photons with a small error, whose photons then produce pairs with a small error of their own, may end up with a final spectrum whose error differs from the single-process estimates;
the errors could compound or cancel.
We leave that question to the first production runs that use these approximations.
For the same reason we quote no speedup:
the cost of an evaluation depends on the host code, the hardware, and the tabulation it replaces, and we have not measured it inside a production particle-in-cell code.

Our approximations are surprisingly accurate even though only low-order polynomials are used.
The regions of largest relative error are typically those where the approximated functions take vanishingly small values, so the absolute error there stays small as well.
These regions are also where particle production is least frequent, and therefore where accuracy matters least.

The solutions for $\chi_\gamma$ and $\chi_e$ diverge at $\zeta=1$, that is, at $r=1$.
The divergence follows from the $\chi$-parameters being inverse functions of the cumulative probabilities, whose derivatives vanish at $r=1$, $\partial_rp(\chi, r=1)=0$, so that
\begin{equation}
  \frac{d\chi_1(\chi, \zeta)}{d\zeta}\bigg|_{\zeta=1}=(\partial_rp(\chi, r
                                                     =1))^{-1}
                                                     =\infty.
\end{equation}
The same divergence appears in the $\xi$-parameters of the spectra in Sect.~\ref{sect:theory}, where $\lim_{r\to 1} \xi = \infty$ for both processes.
Physically, the incoming particle cannot pass all of its energy to one of the produced particles:
both always take a non-zero share.
A numerical implementation must respect this, because random numbers too close to $1$ produce complex values or floating-point errors.
Clipping the generated random numbers,
$\zeta_{\mathrm{max}}=1-\epsilon,$ and $\zeta_\mathrm{min} =\epsilon$ where $\epsilon \ll1$, avoids the problem;
$\epsilon=5\cdot10^{-4}$ for synchrotron radiation and $\epsilon=10^{-4}$ for Breit-Wheeler suffices in our calculations.

The approximations also have a region of applicability.
We optimized synchrotron values, i.e., $T_\mathrm{S}$ and $p_\mathrm{S}$, for $\chi_e\sim(10^{-3}, 10^{2})$.
Below that range the radiation is essentially classical, and above it photon emission becomes rare while other processes take over;
in both directions the functions approach the asymptotic limits that the approximations already carry.
Similarly, we optimized Breit-Wheeler values, i.e., $T_\mathrm{BW}$ and $p_\mathrm{BW}$, for $\chi_\gamma\sim (0.5, 5000)$, below which pair production is exponentially suppressed and above which the same asymptotic argument applies.

The presented method is general and can be easily applied to other processes.
The approach extends to the remaining two first-order QED processes, the $2$-to-$1$ processes of photon absorption by a fermion and one-photon pair annihilation.
Their spectra follow from crossing symmetry once the spin and polarization sums are adjusted for the different incoming and outgoing particles.
A second extension would treat these processes in different vector potentials, e.g., $\mathcal{A}_\mu=(0,0,xB,0)$,
which corresponds to the fields $\mathbf{E}= -\partial_t\mathbf{A}=0$ and $\mathbf{B}= \nabla\times\mathbf{A}=B\hat{z}$, instead of the constant crossed field of Eqs.~\eqref{eq:background_field}--\eqref{eq:const_b} used here.
Although the algorithm is derived for process rates of constant crossed fields, it can still be used in uniform pure magnetic by replacing $\chi$ with the definition $\bar{\chi} = \hbar e |p_\mu G^{\mu \nu}| / m_e^2 c^3$, where $G^{\mu \nu} = \varepsilon^{\mu \nu \alpha \beta } F_{\alpha \beta} /2$ is the Hodge dual of the electromagnetic tensor \cite{nattila2026pair}. The replacement yields accurate results provided that $\bar{\chi}\lesssim 1$.
This makes the presented algorithm also applicable to astrophysical plasma simulations \cite{nattila2024radiative,nattila2026pair, Salmi2026}.

\section*{Acknowledgments}
V.S. and J.N. acknowledge useful discussions with O. Kiuru.
The authors used Anthropic Claude (Opus-5 and Fable-5) models for assistance with manuscript editing and equation checking.

This work is supported by an ERC grant (ILLUMINATOR, 101114623) and by the Research Council of Finland Centre of Excellence in Neutron-Star Physics (project 374063).
The views and opinions expressed are however those of the authors only and do not necessarily reflect those of the European Union or the European Research Council.
Neither the European Union nor the granting authority can be held responsible for them.

\appendix

\section{Light-front coordinates\label{app:lfcoord}}

For a general four-vector with components \cite{seipt2017volkovstatesnonlinearcompton}
\begin{equation}
  a^\mu = (a^0, a^1, a^2, a^3),
\end{equation}
the light-front coordinates are
\begin{equation}
  a^\pm\equiv a^0 \pm a^3,
\end{equation}
\begin{equation}
  \mathbf{a}^\perp=(a^1,a^2),
\end{equation}
and the covariant components are
\begin{equation}
  a_{\mp}=\frac{1}{2}a^{\pm},~\text{and}~\mathbf{a}_\perp
         =-\mathbf{a}^\perp.
\end{equation}
The metric tensor is
\begin{equation}
    g_{\mu \nu}=
    \begin{pmatrix}
        0&\frac{1}{2}&0&0\\
        \frac{1}{2}&0&0&0\\
        0&0&-1&0\\
        0&0&0&-1
    \end{pmatrix},
\end{equation}
and the inverse metric tensor is
\begin{equation}
    g^{\mu \nu}=
    \begin{pmatrix}
        0&2&0&0\\
        2&0&0&0\\
        0&0&-1&0\\
        0&0&0&-1
    \end{pmatrix}.
\end{equation}
The dot product of two light-front four-vectors is
\begin{equation}
  a\cdot b= a^+b_++a^-b_-+\mathbf{a}^\perp \cdot \mathbf{b}_\perp
          =\frac{1}{2}a^+b^-+\frac{1}{2}a^-b^+-\mathbf{a}_\perp \cdot \mathbf{b}_\perp.
\end{equation}
In terms of the metric determinant
\begin{equation}
  \sqrt{-g}=\sqrt{-\det(g_{\mu \nu})}
           =\frac{1}{2},
\end{equation}
the Lorentz-invariant integration measure in light-front coordinates is
\begin{equation}
  \sqrt{-g}~d^4x=\frac{1}{2}dx^+dx^-d^2\mathbf{x}_\perp,
\end{equation}
so that integration over Minkowski space transforms as
\begin{equation}
  \int d^4x=\frac{1}{2}\int dx^+ dx^- d^2 \mathbf{x}^\perp.
\end{equation}
The background field of Eq.~\eqref{eq:background_field} propagates in the negative $z$-direction, which lets us write $x^+$ in terms of the phase,
\begin{equation}
\begin{split}
    \phi = k\cdot x &= \omega (1,0,0,-1)\cdot(x^0, x^1, x^2, x^3)=\omega (x^0 + x^3) \\&= \omega x^+,
\end{split}
\end{equation}
and rewrite the spatial integration in terms of the field phase,
\begin{equation}
  \int d^4x=\frac{1}{2\omega}\int d\phi dx^- d^2 \mathbf{x}^\perp.
\end{equation}
The delta function in light-front coordinates is
\begin{equation}
  \delta^{(4)}(x)=2\delta(x^+)\delta(x^-)\delta^{(2)}(\mathbf{x}^\perp).
\end{equation}
For on-shell particles, which obey the energy-momentum relation of special relativity, the four- and three-dimensional momentum integrals are related by
\begin{equation}
  \int \frac{d^4p}{(2\pi)^4}2\pi\delta(p^2-m^2)=\frac{dp^+ d\mathbf{p}^\perp}{(2\pi)^3 2p^+}
                                               = \frac{dp^+ dp^1 dp^2}{(2\pi)^3 2p^+},
\end{equation}
and the Lorentz-invariant integration measure transforms from Cartesian to light-front coordinates as
\begin{equation}
  \frac{dp^+ dp^1 dp^2}{(2\pi)^3 2p^+} = \frac{dp^1dp^2dp^3}{(2\pi)^3 2p^0}.
\end{equation}

\section{The classical limit of the synchrotron auxiliary function}\label{app:classical_limit}

Numerically integrating the synchrotron spectrum to obtain the cumulative probability of Eq.~\eqref{eq:cumulative_p} runs into floating-point errors in the classical regime $\chi_e \ll1$ and $\chi_\gamma \ll1$, typically below $\chi_e\simeq10^{-4}$.
This appendix derives the limiting value of $T_\mathrm{S}$ analytically, which anchors the low-energy end of the approximation of Eq.~\eqref{eq:t_sync_approx}.

In the classical regime the $\mathcal{O}(\chi_\gamma \xi)$ term vanishes.
For $\chi_\gamma \ll \chi_e$ the term approaches
\begin{equation}
  \frac{3}{2}\chi_\gamma\xi K_{2/3}(\xi) \simeq\frac{3^{2/3}\Gamma(2/3)}{2}\left(\frac{\chi_\gamma^2}{\chi_e}\right)^{2/3},
\end{equation}
which vanishes with decreasing $\chi_\gamma$.
For $\chi_\gamma\lesssim\chi_e$ the classical regime gives $\xi\to \infty$, and the term approaches
\begin{equation}
  \frac{3}{2}\chi_\gamma \xi K_{2/3}(\xi)\simeq \sqrt{\frac{9\pi}{8}}\chi_\gamma \xi^{1/2}e^{-\xi},
\end{equation}
which vanishes exponentially with increasing $\xi$.
The synchrotron spectrum in the classical regime is therefore
\begin{equation}
  \doublederiv{N_\mathrm{S}}{\chi_\gamma}{t} \simeq \frac{\alpha_f m_e ^2 c^4}{\pi \sqrt{3}\hbar \varepsilon_e \chi_e}\left[-\int_{\xi}^{\infty}K_{1/3}(s)\,ds+2 K_{2/3}(\xi)\right].
\end{equation}
The identity \cite{book}
\begin{equation}
  \int_x^\infty K_{5/3}(s)ds = -\int_x ^\infty K_{1/3}(s)ds + 2K_{2/3}(x)
\end{equation}
reduces the bracket to the form familiar from the literature \cite{erber1966high, schwinger1949classical},
\begin{equation}
  \doublederiv{N_\mathrm{S}}{\chi_\gamma}{t} \simeq \frac{\alpha_f m_e ^2 c^4}{\pi \sqrt{3}\hbar \varepsilon_e \chi_e}\int_{\xi}^{\infty}K_{5/3}(s)ds.
\end{equation}
We drop the constant prefactor from here on.
Writing the Bessel function as
\begin{equation}
  K_{5/3}(s)=\int_0 ^\infty \cosh(5t/3)e^{-s\cosh(t)}dt
\end{equation}
turns the spectrum into
\begin{equation}
  \doublederiv{N_\mathrm{S}}{\chi_\gamma}{t} = \frac{1}{\pi \sqrt{3} \chi_e}\int_{0}^{\infty}dt \cosh(5t/3)\int_\xi ^\infty e^{-s\cosh(t)}ds.
\end{equation}
Since $\cosh(t)>0$, the inner integral converges and
\begin{equation}
  \doublederiv{N_\mathrm{S}}{\chi_\gamma}{t} = \frac{1}{\pi \sqrt{3} \chi_e}\int_{0}^{\infty} \frac{\cosh(5t/3)}{\cosh(t)} e^{-\xi\cosh(t)}dt.
\end{equation}
The MC algorithm integrates this spectrum from $\chi_\gamma = 0$ to $\chi_\gamma = r\chi_e$ with $r\in(0,1)$ to obtain the cumulative probability and the auxiliary function, which defines the cumulative production rate
\begin{equation}
  \frac{dN_\mathrm{S}(r)}{dt}=\int_0^{r\chi_e}\doublederiv{N_\mathrm{S}}{\chi'_\gamma}{t} d\chi'_\gamma.
\end{equation}
To carry out the integration in $\xi$, we first substitute
\begin{equation}
  \chi_\gamma=r\chi_e
             \Rightarrow d\chi_\gamma
             =\chi_e dr,
\end{equation}
which expresses $\xi$ through the ratio $r$,
\begin{equation}
  \xi = \frac{2\chi_\gamma}{3\chi_e(\chi_e - \chi_\gamma)}
      =\frac{2r}{3\chi_e(1-r)}.
\end{equation}
Inverting for $r$ gives
\begin{equation}
  r = \frac{3\chi_e \xi}{2+3\chi_e \xi},
\end{equation}
with the differential
\begin{equation}
  dr = \frac{6\chi_e}{(2+3\chi_e \xi)^2}d\xi,
\end{equation}
so that the integral becomes
\begin{equation}
\begin{split}
\deriv{N_\mathrm{S}(r)}{t} = \frac{1}{\pi \sqrt{3}}&\int_{0}^{\infty} dt \frac{\cosh(5t/3)}{\cosh(t)}\times \\ &\int_0 ^{\xi}\frac{6\chi_e}{(2+3\chi_e \xi)^2} e^{-\xi\cosh(t)}d\xi.
\end{split}
\end{equation}
In the classical regime $\chi_e\ll1$, so $\chi_e \xi \gtrsim 1$ requires $\xi \gg1$, where the exponential damps the integrand.
Neglecting the $\chi_e\xi$ term in the denominator leaves
\begin{equation}
  \deriv{N_\mathrm{S}(r)}{t} \simeq\frac{3\chi_e}{2\pi \sqrt{3}}\int_{0}^{\infty} dt \frac{\cosh(5t/3)}{\cosh(t)} \int_0 ^{\xi} e^{-\xi\cosh(t)}d\xi,
\end{equation}
and integrating the exponential gives
\begin{equation}
  \deriv{N_\mathrm{S}(r)}{t} \simeq\frac{3\chi_e}{2\pi \sqrt{3}}\int_{0}^{\infty} dt \frac{\cosh(5t/3)}{\cosh^2(t)} (1- e^{-\xi\cosh(t)}).
\end{equation}
The first of the two resulting integrals is a standard one,
\begin{equation}
  \int_0^\infty \frac{\cosh(\nu t)}{\cosh^2(t)}\,dt = \frac{\pi \nu/2}{\sin(\pi \nu/2)}, \quad |\nu|
                                                    <2,
\end{equation}
which converges here because $\nu=5/3$ lies below the threshold $\nu=2$ at which the integrand stops decaying.
Evaluating it at $\nu=5/3$,
\begin{equation}
  \int_0^\infty dt \frac{\cosh(5t/3)}{\cosh^2(t)}=\frac{5\pi/6}{\sin(5\pi/6)}
                                                 =\frac{5\pi}{3},
\end{equation}
leaves
\begin{equation}
  \deriv{N_\mathrm{S}(r)}{t} \simeq\frac{3\chi_e}{2\pi \sqrt{3}} \left(\frac{5\pi}{3}- \int_{0}^{\infty} dt \frac{\cosh(5t/3)}{\cosh^2(t)}e^{-\xi\cosh(t)}\right).
\end{equation}
As the electron quantum parameter approaches zero over the integration range $r\in (0,1)$,
\begin{equation}
  \lim_{\chi_e\to0} \lim_{r\to 1}\xi=\lim_{\chi_e\to 0}\lim_{r\to 1}\frac{2\chi_\gamma}{3\chi_e(\chi_e-\chi_\gamma)}
                                    = \infty,
\end{equation}
so the remaining integral vanishes.
With Eq.~\eqref{eq:rate_sync}, the leading-order behavior of the auxiliary function at $\chi_e\to 0$ is
\begin{equation}
  \lim_{\chi_e\to0}T_\mathrm{S}(\chi_e)=\lim_{\chi_e\to0}\frac{1}{\chi_e}\deriv{N_\mathrm{S}(r)}{t}\bigg|_{r=1}
                                       =\frac{5}{2\sqrt{3}}
                                       \approx 1.4434.
\end{equation}

\section{Solution of a quartic equation} \label{app:quartic}

Consider a quartic equation \cite{escofier2000galois, tignol2015galois}
\begin{equation} \label{eq:quartic}
  x^4 + Ax^2 + Bx+C=0.
\end{equation}
Its solution starts from the associated resolvent cubic equation
\begin{equation} \label{eq:cubic}
  2y^3-Ay^2 - 2Cy + AC - \frac{B^2}{4}=0,
\end{equation}
which factorizes as
\begin{equation}
  \left(y^2-C\right)\left(2y-A\right)-\frac{B^2}{4}=0,
\end{equation}
and yields
\begin{equation} \label{eq:y_squared}
  y^2 = \frac{B^2}{4\left(2y-A\right)}+C.
\end{equation}
Cardano's method, for instance, solves Eq.~\eqref{eq:cubic} for $y$.
A second-order equation for $x$ then follows from the square
\begin{equation}
  (x^2+y)^2=x^4+2x^2y+y^2,
\end{equation}
in which the $x^4$ term is rewritten with Eq.~\eqref{eq:quartic} and the $y^2$ term with Eq.~\eqref{eq:y_squared},
\begin{equation}
  (x^2 + y)^2=(2y-A)x^2 - Bx + \frac{B^2}{4(2y-A)}.
\end{equation}
The right-hand side is a perfect square,
\begin{equation}
  (x^2 + y)^2=\left(\sqrt{2y-A}x-\frac{B}{2\sqrt{2y-A}}\right)^2,
\end{equation}
so writing the difference of the two squares as a product gives
\begin{equation}
\begin{split}
    &\left(x^2+y-\sqrt{2y-A}x+\frac{B}{2\sqrt{2y-A}}\right)\times \\ &\left(x^2+y+\sqrt{2y-A}x-\frac{B}{2\sqrt{2y-A}}\right)=0.
\end{split}
\end{equation}
One of the two factors must vanish, which gives the four roots
\begin{align}
  x_1^\pm &=\frac{\sqrt{2y-A}\pm\sqrt{-2y-A-\frac{2B}{\sqrt{2y-A}}}}{2}, \\
  x_2^\pm &=\frac{-\sqrt{2y-A}\pm\sqrt{-2y-A+\frac{2B}{\sqrt{2y-A}}}}{2}.
\end{align}
A quartic equation that is not in depressed form,
\begin{equation}
  ax^4 + bx^3 + cx^2 + dx+e=0,
\end{equation}
is brought to one by the change of variable
\begin{equation}
  x=u-\frac{b}{4a},
\end{equation}
which yields
\begin{equation}
  u^4+Ku^2+Mu+N=0,
\end{equation}
with coefficients
\begin{align}
  K &=\frac{c}{a}-\frac{3b^2}{8a^2}, \\
  M &=\frac{b^3}{8a^3}-\frac{bc}{2a^2}+\frac{d}{a}, \\
  N &=\frac{e}{a}-\frac{bd}{4a^2}+\frac{cb^2}{16a^3}-\frac{3b^4}{256a^4}.
\end{align}
The method above then applies.

\subsection{Synchrotron cumulative probability} \label{App:QuarticSyncEq}

Rearranging Eq. \eqref{eq:sync_randomzeta} gives the quartic equation
\begin{equation} \label{eq:sync_quartic_x}
  x^4-\frac{c(1-\zeta)}{a}x^3-\frac{b(1-\zeta)}{a}x^2-\frac{1-\zeta}{a}=0,
\end{equation}
which we solve with the method of App.~\ref{app:quartic}.
Introducing the shorthands $C\equiv -c(1-\zeta)$,
$B\equiv -b(1-\zeta)$,
$D\equiv -(1-\zeta)$,
the change of variable $x=u-C/(4a)$ converts Eq.~\eqref{eq:sync_quartic_x} to the depressed quartic
\begin{equation} \label{eq:sync_quartic_u}
  u^4 + K u^2+M u+N=0,
\end{equation}
with coefficients
\begin{align}
   &K= \frac{B}{a}-\frac{3C^2}{8a^2}, \\
   &M= \frac{C^3}{8a^3}-\frac{BC}{2a^2}, \\
   &N=\frac{D}{a}+\frac{BC^2}{16a^3}-\frac{3C^4}{256a^4}.
\end{align}
The resolvent cubic equation of Eq.~\eqref{eq:sync_quartic_u} is
\begin{equation}
  2y^3 - Ky^2 - 2Ny + KN - \frac{M^2}{4}=0,
\end{equation}
and Cardano's method gives its solution
\begin{equation}
  y = \frac{K}{6}+\sqrt[3]{-\frac{q}{2}+\sqrt{\frac{q^2}{4}+\frac{p^3}{27}}}+\sqrt[3]{-\frac{q}{2}-\sqrt{\frac{q^2}{4}+\frac{p^3}{27}}},
\end{equation}
where
\begin{equation}
  p=\frac{-12N-K^2}{12},
\end{equation}
\begin{equation}
  q=\frac{-2K^3+72KN-27M^2}{216}.
\end{equation}
The four roots of the quartic equation are then \cite{escofier2000galois, tignol2015galois}
\begin{align}
  u_1^\pm &=\frac{\sqrt{2y-K}\pm \sqrt{-2y-K-\frac{2M}{\sqrt{2y-K}}}}{2}, \\
  u_2^\pm &= \frac{-\sqrt{2y-K}\pm \sqrt{-2y-K+\frac{2M}{\sqrt{2y-K}}}}{2}.
\end{align}

The physical solutions come from $u_1^+$ and $u_2^+$, and the sign of $\Delta = -2y - K- 2M/\sqrt{2y-K}$ selects between them:
$u_1^+$ for $\Delta \geq 0$ and $u_2^+$ for $\Delta < 0$.
At the crossover the quartic equation has a double root, so the two expressions agree there and the choice never introduces a discontinuity.
With the appropriate root, the energy fraction is
\begin{equation} \label{eq:sync_r}
  r = \left[1-\left(u-\frac{C}{4a}\right)^{1/n}\right]^3 \, .
\end{equation}

For some rare values of $\chi_e$ and $\zeta$ the term $(2y-K)$ inside the square root can be a small negative number, $\sim -10^{-17}$. In this case, we approximate the solutions with a simpler Padé approximation of the form
\begin{equation}
    p_\mathrm{S}(\chi_e, r) = \frac{(1+g(\chi_e))r^{h(\chi_e)}}{1+g(\chi_e)r^{h(\chi_e)}},
\end{equation}
where the auxiliary parameters are
\begin{equation}
    h(\chi_e) = \frac{0.2645654 + 1.426191\chi_e ^{-0.6562}}{1.166 + 1.261\chi_e ^{-0.6562}},
\end{equation}
\begin{equation}
    g(\chi_e) = \frac{-0.000559811 + 16.49\chi_e ^{-2.67}}{0.00153794 + 3.409\chi_e ^{-1.409}}f(\chi_e),
\end{equation}
\begin{equation}
    f(\chi_e) = 0.965 + 0.7328\chi_e^{0.8064}\exp{\left(-\frac{\chi_e^{0.5187}}{2.74} \right)}.
\end{equation}
These yield a fallback solution
\begin{equation}
    r = \left(\frac{\zeta}{1 + g(\chi_e)(1-\zeta)}\right)^{1/h(\chi_e)}
\end{equation}
This approximation is not as accurate as the quartic Padé model, so it is only used when the full solution does not return a valid value for $r$.

\subsection{Breit Wheeler cumulative probability} \label{app:QuarticBWEq}

Rearranging Eq. \eqref{eq:bw_randomzeta} gives the quartic equation
\begin{equation}
  u^4-\frac{c(1-\zeta)}{a}u^2- \frac{b(1-\zeta)}{a} u-\frac{1-\zeta}{a}=0,
\end{equation}
and which we solve with the method of App.~\ref{app:quartic}.
The corresponding resolvent cubic equation is
\begin{equation}
  2y^3+\frac{c(1-\zeta)}{a}y^2+\frac{2(1-\zeta)}{a}y+\frac{(1-\zeta)^2}{a^2}\left(c-\frac{b^2}{4}\right)=0,
\end{equation}
and Cardano's method yields
\begin{equation}
  y = \sqrt[3]{-\frac{q}{2}+\sqrt{\Delta}}+\sqrt[3]{-\frac{q}{2}-\sqrt{\Delta}}-\frac{c(1-\zeta)}{6a},
\end{equation}
where
\begin{align}
  p      &=\frac{1-\zeta}{12a}\left(12-\frac{(1-\zeta)c^2}{a}\right), \\
  q      &=\frac{(1-\zeta)^2}{108a^2}\left(\frac{(1-\zeta)c^3}{a}+36c-\frac{27b^2}{2}\right), \\
  \Delta &=\frac{q^2}{4}+\frac{p^3}{27}.
\end{align}
The roots of the quartic equation are \cite{escofier2000galois, tignol2015galois}
\begin{align}
  u_1^{\pm}=\frac{1}{2}\bigg( &\sqrt{2y+c(1-\zeta)/a} \\
                              &\pm \sqrt{-2y+\frac{c(1-\zeta)}{a}+\frac{2b(1-\zeta)}{a\sqrt{2y+c(1-\zeta)/a}}}\bigg), \\
  u_2^{\pm}=\frac{1}{2}\bigg( &-\sqrt{2y+c(1-\zeta)/a} \\
                              &\pm \sqrt{-2y+\frac{c(1-\zeta)}{a}-\frac{2b(1-\zeta)}{a\sqrt{2y+c(1-\zeta)/a}}}\bigg).
\end{align}
Only $u_2^+$ gives real positive solutions, so the energy fraction is

\begin{equation} \label{eq:bw_r}
    r(\chi_\gamma, \zeta)=
    \begin{cases}
        1-[u_2^+(\chi_\gamma, \zeta)]^{1/n},~&\text{for} ~\zeta\geq0.5\\
        [u_2^+(\chi_\gamma, 1-\zeta)]^{1/n},~&\text{for} ~\zeta<0.5,
    \end{cases}
\end{equation}
The form for $\zeta < 0.5$ follows from manual symmetrization.

\section{Auxiliary parameters}

This appendix describes how we developed the parameter functions of the cumulative probabilities.

\subsection{Synchrotron radiation} \label{app:param_sync}

\begin{figure*}[h]
    \centering
    \includegraphics[scale=0.625, clip=true, trim={0.0cm 0 0 0}]{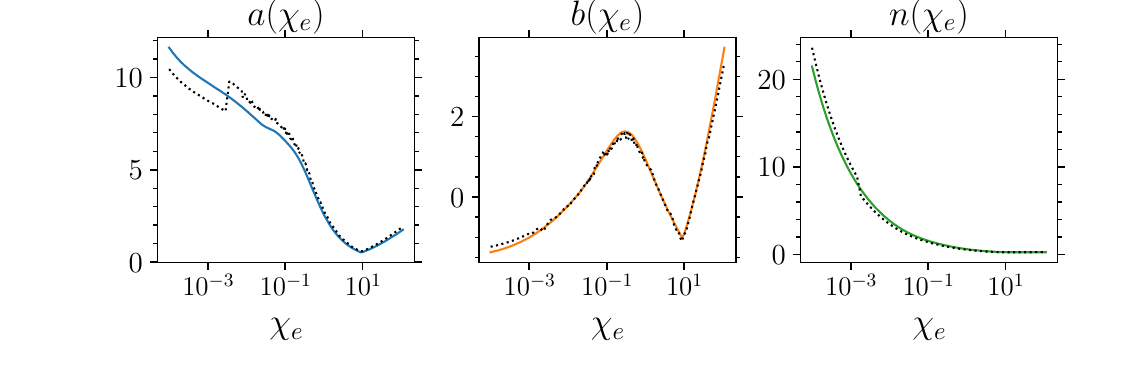}
    \includegraphics[scale=0.55, clip=true]{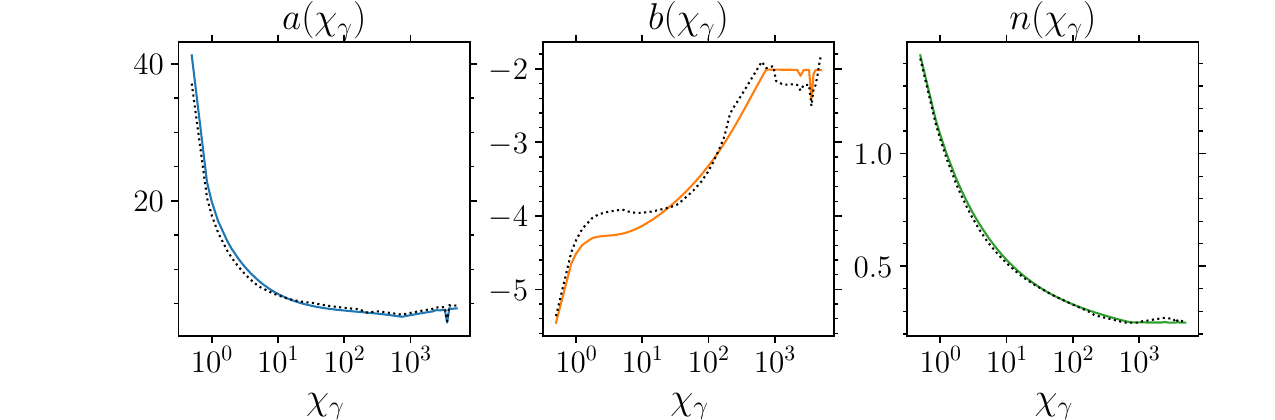}
    \caption{Top row: Optimal values of the auxiliary parameters in the synchrotron Padé approximant obtained by fitting the probability (solid curves) and the power spectrum (dashed curves). The weight in the error sum is $\sigma_i=1/y(\chi_e,\chi_{\gamma,i})^{\exp(-\chi_e/20)}$, where $y(\chi_e,\chi_{\gamma,i})$ are the power spectrum values; the exponential removes the weight at large $\chi_e$ where the sharp peak around $r\approx1$ would have a large weight in the error minimization and cause overall inaccuracies. Bottom row: Optimal values of the auxiliary parameters in the Breit-Wheeler Padé approximant obtained by fitting the probability (solid curves) and the spectrum (dashed black curves). The weight in the error sum is $\sigma_i = 1/(y_i(\chi_\gamma, \chi_{e, i})\chi_{e, i})$, where $y(\chi_\gamma, \chi_e)$ is the approximated function itself.}
    \label{fig:sync_BW_abn_params}
\end{figure*}

\begin{figure*}[ht]
    \centering
    \includegraphics[scale=0.48, clip=true, trim={0cm 0.2cm 0.0cm 0.125cm}]{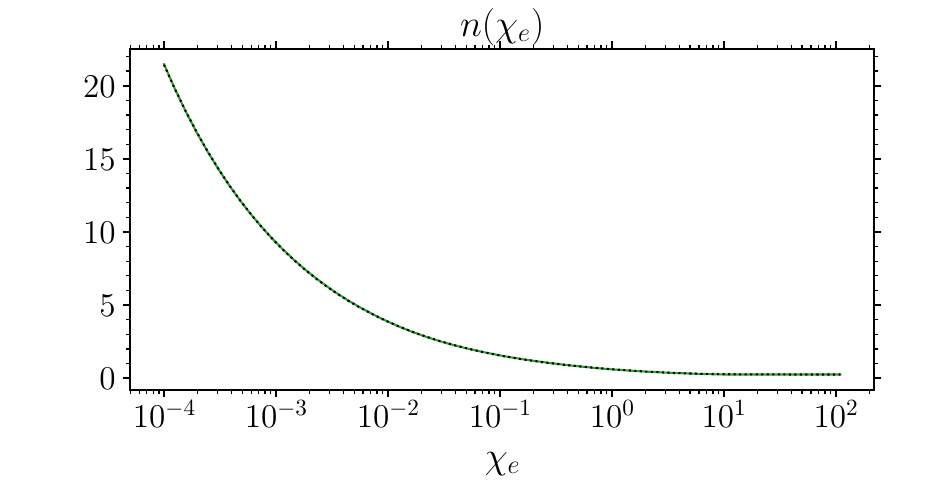}
     \includegraphics[scale=0.48, clip=true, trim={0cm 0.2cm 0.0cm 0.1cm}]{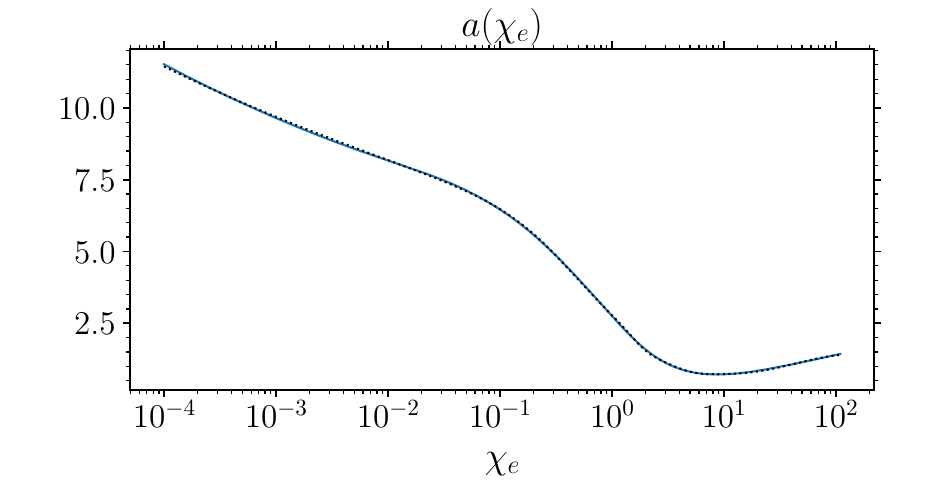}
      \includegraphics[scale=0.48, clip=true, trim={0cm 0.2cm 0.0cm 0.15cm}]{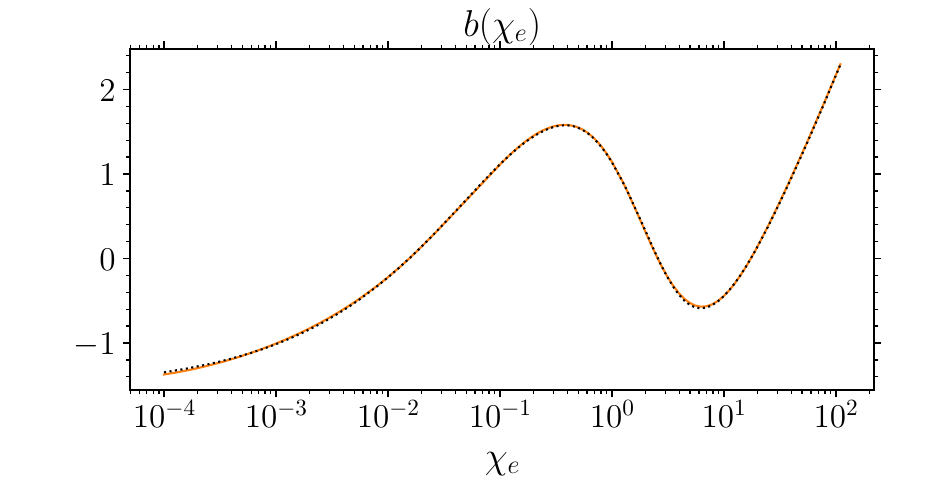}
    \caption{Approximating forms for synchrotron processes'  auxiliary parameters $n(\chi_e)$ (Eq. \ref{eq:sync_n}), $a(\chi_e)$ (Eq. \ref{eq:sync_a}), and $b(\chi_e)$ (Eq. \ref{eq:sync_b}) in dotted lines compared to the values obtained by fitting the approximated probability to the numerical values in solid lines.}
    \label{fig:sync_abn_approximations}
\end{figure*}

We obtain the optimal values for auxiliary parameters $n(\chi_e)$, $a(\chi_e)$, and $b(\chi_e)$ by fitting Eqs.~\eqref{eq:sync_pade} and \eqref{eq:sync_p_deriv} to the numerical probability and radiation spectrum through the error sum of Eq.~\eqref{eq:err_sum}.

For the optimal values of $n(\chi_e)$, we found the following approximating form
\begin{equation} \label{eq:sync_n}
\begin{split}
n(\chi_e)=\frac{\lambda_0\chi_e^{-k_0}}{1+\rho_0\chi_e^{k_0}+\rho_1\chi_e^{\ell_0}} + \frac{1}{4}.
\end{split}
\end{equation}
Fixing $n(\chi_e)$ to this form, the optimal values of $a(\chi_e)$ follow
\begin{equation} \label{eq:sync_a}
\begin{split}
    a(\chi_e)=\bigg(&C_a^0+\frac{\lambda_1\chi_e^{k_1}}{1+\rho_2\chi_e^{\ell_1}} C_{a}^1(\chi_e)+ \frac{\lambda_2\chi_e^{k_2}}{1+\rho_3\chi_e^{k_2}}\bigg)C_a^2(\chi_e),
\end{split}
\end{equation}
where the Gaussian correction functions are
\begin{align}
  C_a^1(\chi_e) &=1+\alpha_2e^{-\Delta_2(\chi_e-\delta_2)^2}, \label{eq:sync_ca1} \\
  C_a^2(\chi_e) &=1+\alpha_3e^{-\Delta_3(\ln(\chi_e)-\delta_3)^2}. \label{eq:sync_ca2}
\end{align}
The optimal values of $b(\chi_e)$ follow
\begin{equation} \label{eq:sync_b}
\begin{split}
    b(\chi_e)=\bigg(&C_b^0+\left(\beta_0\chi_e^{k_3}+\beta_1\chi_e^{k_4}\right)C_b^1(\chi_e)  \\& +\frac{\lambda_3\chi_e^{k_5}+\lambda_4\chi_e^{k_6}}{1+\rho_4\chi_e^{k_5}}\bigg)C_b^2(\chi_e),
\end{split}
\end{equation}
with the correction functions
\begin{align}
  C_b^1(\chi_e) &=\exp\left({-\Delta_4\chi_e^{k_7}}\right), \label{eq:sync_cb1} \\
  C_b^2(\chi_e) &=\bigg(\alpha_4 + \alpha_5\exp\left({-\Delta_5\left(\ln(\chi_e)+\delta_4\right)^2}\right)\bigg)^{k_8}. \label{eq:sync_cb2}
\end{align}
Table~\ref{tab:sync_params} lists the values of the symbols in Eqs.~\eqref{eq:sync_n}--\eqref{eq:sync_cb2}.

\begin{table}
\centering
\begin{tabular}{lllllllll}
     & $n(\chi_e)$ &  &  & $a(\chi_e)$ &  &  & $b(\chi_e)$ &  \\ \hline
     Eq. & Var. & Value & Eq. & Var. & Value & Eq. & Var. & Value \\
     \hline
     & $\lambda_0$ & 0.92674 &  & $C_a^0$ & 0.53700 &  & $C_b^0$ & -1.56564 \\

     & $\rho_0$ & $1.26886$ &  & $\lambda_1$ & 5.37351 &  & $\beta_0$ & 11.3943 \\

    \eqref{eq:sync_n} & $\rho_1$ & 0.30839 &  & $\rho_2$ & 1.74095 &  & $k_3$ & 0.43384 \\

     & $k_0$ & 0.34532 & \eqref{eq:sync_a} & $k_1$ & -0.07621 &  & $\beta_1$ & 8.79281 \\

     & $\ell_0$ & 1.74193 &  & $\ell_1$ & 1.29055 & \eqref{eq:sync_b} & $k_4$ & 0.88136 \\
     \cline{1-3}
     &  &  &  & $\lambda_2$ & 0.00150 &  & $\lambda_3$ & -0.38800 \\

     &  &  &  & $\rho_3$ & 0.00150 &  & $k_5$ & 2.23683 \\

     &  &  &  & $k_2$ & 1.77253 &  & $\lambda_4$ & 0.32931 \\ \cline{4-6}

     &  &  &  & $\alpha_2$ & 0.1648 &  & $k_6$ & 2.35484 \\

     &  &  & \eqref{eq:sync_ca1} & $\delta_2$ & 1.2024 &  & $\rho_4$ & 0.04777 \\ \cline{7-9}

     &  &  &  & $\Delta_2$ & 2.279 & \eqref{eq:sync_cb1} & $\Delta_4$ & 1.99971 \\ \cline{4-6}

     &  &  &  & $\alpha_3$ & -0.08494 &  & $k_7$ & 0.46344 \\ \cline{7-9}

     &  &  & \eqref{eq:sync_ca2} & $\delta_3$ & 1.75725 &  & $\alpha_4$ & 0.87786 \\

     &  &  &  & $\Delta_3$ & 4.59297 &  & $\alpha_5$ & 0.65635 \\ \cline{4-6}

     &  &  &  &  &  & \eqref{eq:sync_cb2} & $\Delta_5$ & 1.44269 \\

     &  &  &  &  &  &  & $\delta_4$ & 0.27729 \\

     &  &  &  &  &  &  & $k_8$ & 0.06620 \\  \cline{7-9}

\end{tabular}
\caption{Values of the symbols in Eqs.~\eqref{eq:sync_n}--\eqref{eq:sync_cb2}, which give the parameters $n(\chi_e)$, $a(\chi_e)$, and $b(\chi_e)$ of the synchrotron cumulative probability.}
\label{tab:sync_params}
\end{table}

Leaving $a(\chi_e)$, $b(\chi_e)$, and $n(\chi_e)$ all free produces noisy optimal values and a very sudden change in the sign of the derivative around $\chi_e\approx10$.
The flexibility of the approximant is what causes this:
many different parameter combinations give an equally good fit.
To remove the noise and the discontinuities, we fix an approximate form for one parameter and re-fit the other two with that form held fixed.
We treat $n(\chi_e)$ first because it is the simplest and behaves the most smoothly, as Fig.~\ref{fig:sync_BW_abn_params} shows.
The same figure shows that fitting the approximated spectrum to the numerical values gives less stable parameters than fitting the approximated probability, so we take the parameter values from the probability fit.
We nevertheless use the power spectrum $dP_\mathrm{rad}/d\chi_\gamma$ as the weight in that fit, which emphasizes the regions carrying most of the radiated energy.
Figure \ref{fig:sync_abn_approximations} compares the resulting approximating forms with the fitted values.

\subsection{Breit-Wheeler process} \label{app:param_bw}

For the optimal values of $n(\chi_\gamma)$, we found the following approximating form
\begin{equation} \label{eq:bw_n}
\begin{split}
    n(\chi_\gamma)= \frac{1}{4}&+\frac{\theta_0}{1+\phi_0((\frac{\chi_\gamma}{100})^{\omega_0}+(\frac{\chi_\gamma}{100})^{2\omega_0})}+\theta_1e^{-\upsilon_0\left(\frac{\chi_\gamma}{1000}\right)^{\varepsilon_0}}.
\end{split}
\end{equation}
Fixing $n(\chi_\gamma)$ to this form, the optimal values of $a(\chi_\gamma)$ follow
\begin{equation}
\label{eq:bw_a}
\begin{split}
    a(\chi_\gamma) = \left(\theta_2\left(\frac{\chi_\gamma}{150}\right)^{\varepsilon_1} +\theta_3\left(\frac{\chi_\gamma}{150}\right)^{\varepsilon_2}-1\right)&\\\times\left(1+\theta_4\left(\frac{\chi_\gamma}{150}\right)^{\varepsilon_3}\right)&,
\end{split}
\end{equation}
and those of $b(\chi_\gamma)$ follow
\begin{equation}
\label{eq:bw_b}
\begin{split}
    b(\chi_\gamma) = \theta_5-\frac{1+\theta_6\left(\frac{\chi_\gamma}{100}\right)^{\varepsilon_4}}{1+\phi_1\left(\frac{\chi_\gamma}{100}\right)^{\omega_1}}+\theta_7\chi_\gamma^{\varepsilon_5}.
\end{split}
\end{equation}
Table~\ref{tab:bw_params} lists the values of the symbols in Eqs.~\eqref{eq:bw_n}--\eqref{eq:bw_b}.

\begin{table*}[b!]
\centering
\begin{tabular}{lllllllll}
     & $n(\chi_\gamma)$ &  &  & $a(\chi_\gamma)$ &  &  & $b(\chi_\gamma)$ &  \\ \hline
     Eq. & Var. & Value & Eq. & Var. & Value & Eq. & Var. & Value \\
     \hline
     & $\theta_0$ & 14.31297 &  & $\theta_2$ & 0.14550 &  & $\theta_5$ & -1.03339 \\
     & $\phi_0$ & 250.28545 &  & $\theta_3$ & 2.10323 &  & $\theta_6$ & 6.74438 \\
   \eqref{eq:bw_n}  & $\omega_0$ & 0.57126 & \eqref{eq:bw_a} & $\theta_4$ & 1.89599 & \eqref{eq:bw_b} & $\varepsilon_4$ & -1.59944 \\
     & $\theta_1$ & 0.14229 &  &  $\varepsilon_1$ & -0.28814 &  & $\phi_1$ & 2.37683\\
     & $\upsilon_0$ & 4.41464 &  & $\varepsilon_2$ & 0.23634 &  & $\omega_1$ & -1.56170 \\
     & $\varepsilon_0$ & 0.63897 &  & $\varepsilon_3$ & -0.74460 &  & $\theta_7$ & -0.13742 \\
     &  &  &  &  &  &  & $\varepsilon_5$ & -2.75390 \\

\end{tabular}
\caption{Values of the symbols in Eqs.~\eqref{eq:bw_n}--\eqref{eq:bw_b}, which give the parameters $n(\chi_\gamma)$, $a(\chi_\gamma)$, and $b(\chi_\gamma)$ of the Breit-Wheeler cumulative probability.}
\label{tab:bw_params}
\end{table*}

\begin{figure*}[ht]
    \centering
    \includegraphics[scale=0.52, clip=true, trim={0cm 0.5cm 0.0cm 0.3cm}]{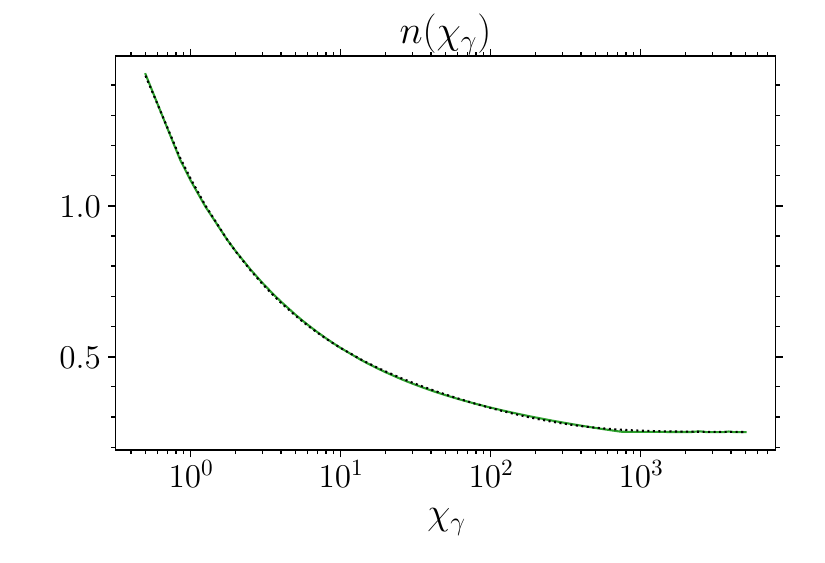}
     \includegraphics[scale=0.52, clip=true, trim={0cm 0.5cm 0.0cm 0.3cm}]{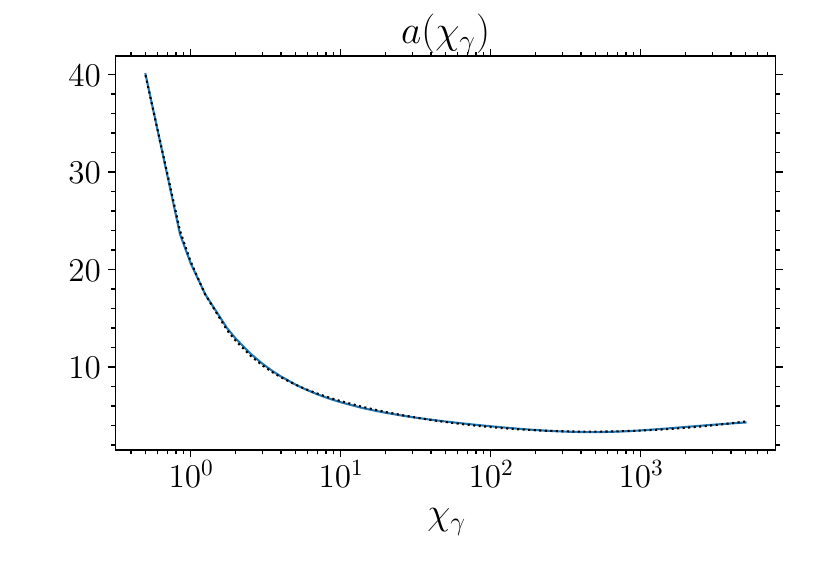}
      \includegraphics[scale=0.52, clip=true, trim={0cm 0.5cm 0.0cm 0.3cm}]{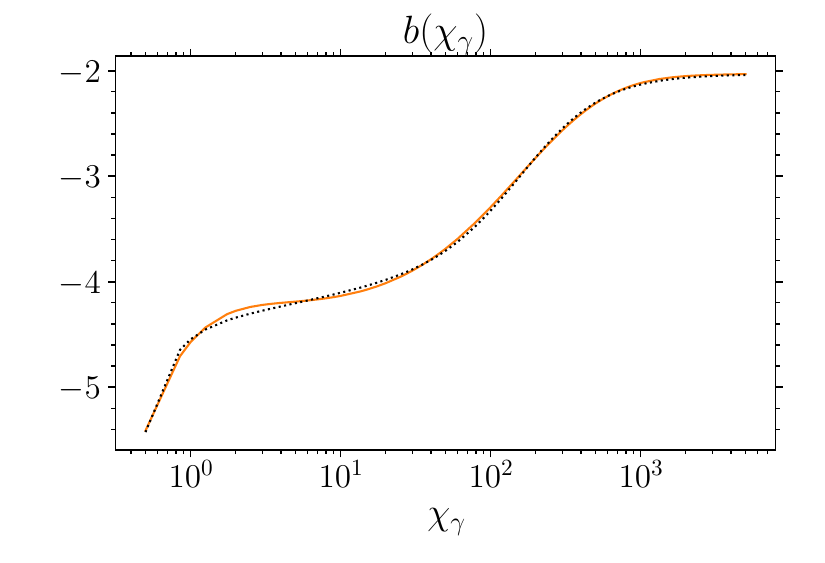}
    \caption{Approximating forms for Breit-Wheeler processes' auxiliary parameters $n(\chi_\gamma)$ (Eq. \ref{eq:bw_n}), $a(\chi_\gamma)$ (Eq. \ref{eq:bw_a}), and $b(\chi_\gamma)$ (Eq. \ref{eq:bw_b}) in dotted lines compared to the values obtained by fitting the approximated probability to the numerical values in solid lines.}
    \label{fig:bw_abn_approximations}
\end{figure*}

The optimal parameter values jump discontinuously around $\chi_\gamma\approx775$.
With increasing $\chi_\gamma$ the spectrum concentrates towards the ends of the interval, and the cheapest way for the fit to follow it is to drive the power parameter to its minimum value $n(\chi_\gamma)\to1/4$, as Eq.~\eqref{eq:bw_p_deriv} shows.
We remove the discontinuity as in the synchrotron case, by fixing a continuous approximating form for $n(\chi_\gamma)$ and re-fitting $a(\chi_\gamma)$ and $b(\chi_\gamma)$ with it.
Figure~\ref{fig:sync_BW_abn_params} shows the optimal values, and Figure \ref{fig:bw_abn_approximations} compares them with the approximating forms.

\clearpage

\bibliography{refs}

@PREAMBLE{
 "\providecommand{\noopsort}[1]{}" 
 # "\providecommand{\singleletter}[1]{#1}%" 
}

@article{lobet2016modeling,
  title={Modeling of radiative and quantum electrodynamics effects in PIC simulations of ultra-relativistic laser-plasma interaction},
  author={Lobet, Mathieu and d'Humi{\`e}res, Emmanuel and Grech, Mickael and Ruyer, Charles and Davoine, Xavier and Gremillet, Laurent},
  journal={Journal of Physics: Conference Series},
  volume={688},
  number={1},
  pages={012058},
  year={2016},
  publisher={IOP Publishing},
  doi = {10.1088/1742-6596/688/1/012058}
}

@phdthesis{martinez2018radiative,
  title={Radiative and quantum electrodynamic effects in ultra-relativistic laser-matter interaction},
  author={Martinez, Bertrand},
  year={2018},
  school={Universit{\'e} de Bordeaux}
}

@article{niel2018quantum,
  title={From quantum to classical modeling of radiation reaction: A focus on stochasticity effects},
  author={Niel, Fabien and Riconda, C and Amiranoff, Fran{\c{c}}ois and Duclous, R and Grech, M},
  journal={Physical Review E},
  volume={97},
  number={4},
  pages={043209},
  year={2018},
  publisher={APS}
}

@misc{seipt2017volkovstatesnonlinearcompton,
      title={Volkov States and Non-linear Compton Scattering in Short and Intense Laser Pulses}, 
      author={Daniel Seipt},
      year={2017},
      eprint={1701.03692},
      archivePrefix={arXiv},
      primaryClass={physics.plasm-ph},
      url={https://arxiv.org/abs/1701.03692}, 
}

@book{berestetskii2012quantum,
  title={Quantum Electrodynamics: Volume 4},
  author={Berestetskii, Vladimir Borisovich and Pitaevskii, Lev Petrovich and Lifshitz, Evgenii Mikhailovich},
  volume={4},
  year={2012},
  publisher={Elsevier}
}

@article{advancesinqed,
  title={Advances in QED with intense background fields},
  author={Fedotov, A and Ilderton, A and Karbstein, F and King, Ben and Seipt, D and Taya, H and Torgrimsson, Greger},
  journal={Physics Reports},
  volume={1010},
  pages={1--138},
  year={2023},
  publisher={Elsevier}
}

@book{peskin2018introduction,
  title={An Introduction to quantum field theory},
  author={Peskin, Michael E},
  year={2018},
  publisher={CRC press}
}

@article{nikishov1964quantum,
  title={Quantum processes in the field of a plane electromagnetic wave and in a constant field. I},
  author={Nikishov, AI and Ritus, VI},
  journal={Sov. Phys. JETP},
  volume={19},
  number={2},
  pages={529--541},
  year={1964}
}

@article{ritus1985quantum,
  title={Quantum effects of the interaction of elementary particles with an intense electromagnetic field},
  author={Ritus, VI},
  journal={J. Sov. Laser Res.;(United States)},
  volume={6},
  number={5},
  year={1985}
}

@article{kirk2014modelling,
  title={Modelling gamma-ray photon emission and pair production in high-intensity laser--matter interactions},
  author={Ridgers, Christopher Paul and Kirk, John G and Duclous, Roland and Blackburn, Thomas G and Brady, Christopher S and Bennett, Keith and Arber, Tony D and Bell, Anthony R},
  journal={Journal of Computational Physics},
  volume={260},
  pages={273--285},
  year={2014},
  publisher={Elsevier}
}

@book{book,
author = {Melrose, Don},
year = {2013},
month = {01},
pages = {},
title = {Quantum Plasmadynamics, Magnetized Plasmas},
volume = {854},
isbn = {978-1-4614-4044-4},
doi = {10.1007/978-1-4614-4045-1}
}

@article{erber1966high,
  title={High-energy electromagnetic conversion processes in intense magnetic fields},
  author={Erber, Thomas},
  journal={Reviews of Modern Physics},
  volume={38},
  number={4},
  pages={626},
  year={1966},
  publisher={APS}
}

@article{schwinger1949classical,
  title={On the classical radiation of accelerated electrons},
  author={Schwinger, Julian},
  journal={Physical review},
  volume={75},
  number={12},
  pages={1912},
  year={1949},
  publisher={APS}
}

@book{abramowitz1948handbook,
  title={Handbook of mathematical functions with formulas, graphs, and mathematical tables},
  author={Abramowitz, Milton and Stegun, Irene A},
  volume={55},
  year={1948},
  publisher={US Government printing office}
}

@book{iterativemethodspade,
  title={Iterative methods for solving nonlinear equations and systems},
  author={Torregrosa, Juan R and Cordero, Alicia and Soleymani, Fazlollah},
  year={2019},
  publisher={MDPI}
}

@article{fedeli2022picsar,
  title={PICSAR-QED: a Monte Carlo module to simulate strong-field quantum electrodynamics in particle-in-cell codes for exascale architectures},
  author={Fedeli, Luca and Zaïm, Neïl and Sainte-Marie, Antonin and Thévenet, Maxence and Huebl, Axel and Myers, Andrew and Vay, Jean-Luc and Vincenti, Henri},
  journal={New Journal of Physics},
  volume={24},
  number={2},
  pages={025009},
  year={2022},
  publisher={IOP Publishing}
}

@book{escofier2000galois,
  title={Galois theory},
  author={Escofier, Jean-Pierre},
  volume={204},
  year={2000},
  publisher={Springer Science \& Business Media}
}

@book{tignol2015galois,
  title={Galois' theory of algebraic equations},
  author={Tignol, Jean-Pierre},
  year={2015},
  publisher={World Scientific Publishing Company}
}

@article{Hofmann:202177,
      author        = "Hofmann, A",
      title         = "{Characteristics of synchrotron radiation}",
      reportNumber  = "CERN-LEP-DI-89-55",
      year          = "1990",
      url           = "https://cds.cern.ch/record/202177",
      doi           = "10.5170/CERN-1990-003.115",
      journal = ""
}

@article{timokhin2010time,
  title={Time-dependent pair cascades in magnetospheres of neutron stars--I. Dynamics of the polar cap cascade with no particle supply from the neutron star surface},
  author={Timokhin, AN},
  journal={Monthly Notices of the Royal Astronomical Society},
  volume={408},
  number={4},
  pages={2092--2114},
  year={2010},
  publisher={Blackwell Publishing Ltd Oxford, UK}
}

@article{grismayer2017seeded,
  title={Seeded QED cascades in counterpropagating laser pulses},
  author={Grismayer, Thomas and Vranic, Marija and Martins, Joana L and Fonseca, RA and Silva, Lu{\'\i}s O},
  journal={Physical Review E},
  volume={95},
  number={2},
  pages={023210},
  year={2017},
  publisher={APS}
}

@article{nerush2011laser,
  title={Laser field absorption in self-generated electron-positron pair plasma},
  author={Nerush, EN and Kostyukov, I Yu and Fedotov, AM and Narozhny, NB and Elkina, NV and Ruhl, H},
  journal={Physical review letters},
  volume={106},
  number={3},
  pages={035001},
  year={2011},
  publisher={APS}
}

@article{nattila2024radiative,
  title={Radiative plasma simulations of black hole accretion flow coronae in the hard and soft states},
  author={N{\"a}ttil{\"a}, Joonas},
  journal={Nature Communications},
  volume={15},
  number={1},
  pages={7026},
  year={2024},
  publisher={Nature Publishing Group UK London}
}

@article{volokitin2023optimized,
  title={Optimized event generator for strong-field QED simulations within the hi-$\chi$ framework},
  author={Volokitin, Valentin and Magnusson, Joel and Bashinov, Aleksei and Efimenko, Evgeny and Muraviev, Alexander and Meyerov, Iosif},
  journal={Journal of Computational Science},
  volume={74},
  pages={102170},
  year={2023},
  publisher={Elsevier}
}

@article{duclous2011monte,
  title={Monte Carlo calculations of pair production in high-intensity laser--plasma interactions},
  author={Duclous, Roland and Kirk, John G and Bell, Anthony R},
  journal={Plasma Physics and Controlled Fusion},
  volume={53},
  number={1},
  pages={015009},
  year={2011}
}

@article{guo2022improving,
  title={Improving the accuracy of hard photon emission by sigmoid sampling of the quantum-electrodynamic table in particle-in-cell Monte Carlo simulations},
  author={Guo, Yinlong and Geng, Xuesong and Ji, Liangliang and Shen, Baifei and Li, Ruxin},
  journal={Physical Review E},
  volume={105},
  number={2},
  pages={025309},
  year={2022},
  publisher={APS}
}

@article{nattila2026pair,
  title={Pair Discharges and Radio Emission from Pulsar Magnetospheres},
  author={N{\"a}ttil{\"a}, Joonas and Salmi, Tuomo},
  journal={arXiv preprint arXiv:2608.21920},
  year={2026}
}

@ARTICLE{Salmi2026,
       author = {{Salmi}, Tuomo and {N{\"a}ttil{\"a}}, Joonas},
        title = {Pair Discharges and Radio Emission from Millisecond-Pulsar and White-Dwarf Magnetospheres},
      journal = {arXiv e-prints},
         year = 2026,
        month = aug,
archivePrefix = {arXiv},
       eprint = {2608.22027},
 primaryClass = {astro-ph.HE},
       adsurl = {https://ui.adsabs.harvard.edu/abs/2026arXiv260822027S}
}

\end{document}